\documentclass[sigconf, authorversion]{acmart}

\AtBeginDocument{%
  \providecommand\BibTeX{{%
    \normalfont B\kern-0.5em{\scshape i\kern-0.25em b}\kern-0.8em\TeX}}}

\usepackage{todonotes}
\usepackage[nolist,nohyperlinks ]{acronym}
\usepackage{siunitx}
\usepackage{calculator}
\usepackage{csquotes}
\usepackage{subcaption} 
\usepackage{fmtcount}
\usepackage{enumitem}
\setlist[description]{%
labelindent=0.5\parindent,%
itemindent=-.6em,%
leftmargin=*,%
}

\begin{acronym}[UMLX]
	\acro{HMD}{head-mounted display}
	\acro{GUI}{graphical user interface}
 	\acro{UI}{user interface}
	\acro{HUD}{head-up display}
	\acro{TCT}{task-completion time}
	\acro{EMM}{estimated marginal mean}
	\acro{AR}{Augmented Reality}
 	\acro{MR}{Mixed Reality}
	\acro{VR}{Virtual Reality}
    \acro{VE}{Virtual Environment}
	\acro{CSCW}{Computer-Supported Cooperative Work}
	\acro{HCI}{Human-Computer Interaction}
	\acro{ABI}{Around-Body Interaction}
	\acro{SLAM}{Simultaneous Location and Mapping}
	\acro{ROM}{range of motion}
	\acro{EMG}{electromyography}
	\acro{STS}{System Trust Scale}
	\acro{IPQ}{Igroup Presence Questionnaire}
	\acro{TLX}{NASA Task Load Index}
 	\acro{RTLX}{Raw Nasa-TLX}
 	\acro{CAD}{Computer Aided Design}
    \acro{ART}{Aligned Rank Transform}
	\acro{IOS}{Inclusion of Other in the Self}
	\acro{GEQ}{Game Experience Questionnaire}
 	\acro{FOV}{Field of View}
    \acro{AI}{Artificial Intelligence}
    \acro{WUI}{Walking User Interface}

\end{acronym}

\copyrightyear{2025}
\acmYear{2025}
\setcopyright{acmlicensed}
\acmConference[CHI '25]{CHI Conference on Human Factors in Computing Systems}{April 26-May 1, 2025}{Yokohama, Japan}
\acmBooktitle{CHI Conference on Human Factors in Computing Systems (CHI '25), April 26-May 1, 2025, Yokohama, Japan}
\acmDOI{10.1145/3706598.3714258}
\acmISBN{979-8-4007-1394-1/25/04}
\newcommand{\systemname}{AR You on Track?}
\newcommand{\longtitle}{Investigating Effects of Augmented Reality Anchoring on Dual-Task Performance While Walking}

\newcommand{\ivPhysTask}{\textsc{Walking Task}}
\newcommand{\ivDifficulty}{\textsc{Virtual Task Difficulty}}
\newcommand{\ivAnchor}{\textsc{Anchoring}}

\newcommand{\pHead}{\texttt{head}}
\newcommand{\pTorso}{\texttt{torso}}
\newcommand{\pHand}{\texttt{hand}}

\newcommand{\NoWalking}{\texttt{No\,Walking}}
\newcommand{\Walking}{\texttt{Walking}}

\newcommand{\conWalkingonly}{\texttt{walking-only}}
\newcommand{\condition}[2]{\texttt{#1-back/#2}}
\newcommand{\nback}[1]{\texttt{#1-back}}

\newcommand{\dvAccuracy}{\texttt{Accuracy}} 
\newcommand{\dvMissedAnswerRate}{\texttt{Missed\,Answer\,Rate}} 
\newcommand{\dvAnswerTime}{\texttt{Answer\,Time}}

\newcommand{\dvWalkingError}{\texttt{Walking\,Error}}
\newcommand{\dvStrideDuration}{\texttt{Stride\,Duration}}
\newcommand{\dvStrideLength}{\texttt{Stride\,Length}}
\newcommand{\dvStrideWidth}{\texttt{Stride\,Width}}
\newcommand{\dvWalkingSpeed}{\texttt{Walking\,Speed}}

\newcommand{\dvRTLX}{\texttt{\acs{RTLX}}}

\newcommand{\ano}[4]{$F_{#1, #2}=#3$, $p#4$}

\newcommand{\subEtaG}[2]{%
	\ifthenelse{\equal{#1}{\string >.05}}
	{}
	{, $\eta_{G}^{2}=#2$}%
}

\newcommand{\subEta}[2]{%
	\ifthenelse{\equal{#1}{\string >.05}}
	{}
	{, $\eta^{2}=#2$}%
}

\newcommand{\chisq}[3]{$\chi^2(#1) = #2$, $p #3$}

\def\ges{$\eta_{G}^{2}$}

\newcommand{\efETAsquared}[1]{%
	\ifdim#1pt>0.139pt 
	large (\ges{} = #1)
	\else 
	\ifdim#1pt>0.059pt 
	medium (\ges{} = #1)
	\else 
	small (\ges{} = #1)
	\fi
	\fi
}

\newcommand{\percentVal}[2]{M=#1\%, SD=#2\%}

\usepackage{etoolbox}

\begin{document}

\title[\systemname{}]{\systemname{} \longtitle{}}

\author{Julian Rasch}
\orcid{0000-0002-9981-6952}
\affiliation{
  \institution{LMU Munich}
  \streetaddress{Frauenlobstr. 7A}
  \city{Munich}
  \country{Germany}
  \postcode{80337}
}
\email{julian.rasch@ifi.lmu.de}

\author{Matthias Wilhalm}
\orcid{0009-0003-0935-6679}
\affiliation{
  \institution{LMU Munich}
  \streetaddress{Frauenlobstr. 7A}
  \city{Munich}
  \country{Germany}
  \postcode{80337}
}
\email{matthias.wilhalm@campus.lmu.de}

\author{Florian Müller}
\orcid{0000-0002-9621-6214}
\affiliation{
  \institution{TU Darmstadt}
  \streetaddress{Hochschulstraße 10}
  \city{Darmstadt}
  \country{Germany}
  \postcode{64289}
}
\email{florian.mueller@tu-darmstadt.de}

\author{Francesco Chiossi}
\orcid{0000-0003-2987-7634}
\affiliation{
  \institution{LMU Munich}
  \streetaddress{Frauenlobstr. 7A}
  \city{Munich}
  \country{Germany}
  \postcode{80337}
}
\email{francesco.chiossi@ifi.lmu.de}

\renewcommand{\shortauthors}{Rasch et al.}


\begin{abstract}

With the increasing spread of AR head-mounted displays suitable for everyday use, interaction with information becomes ubiquitous, even while walking. However, this requires constant shifts of our attention between walking and interacting with virtual information to fulfill both tasks adequately. Accordingly, we as a community need a thorough understanding of the mutual influences of walking and interacting with digital information to design safe yet effective interactions.
Thus, we systematically investigate the effects of different AR anchors (hand, head, torso) and task difficulties on user experience and performance. We engage participants ($n=26$) in a dual-task paradigm involving a visual working memory task while walking. We assess the impact of dual-tasking on both virtual and walking performance, and subjective evaluations of mental and physical load. Our results show that head-anchored AR content least affected walking while allowing for fast and accurate virtual task interaction, while hand-anchored content increased reaction times and workload.

\end{abstract}



\begin{teaserfigure}
	\includegraphics[width=1.0\textwidth]{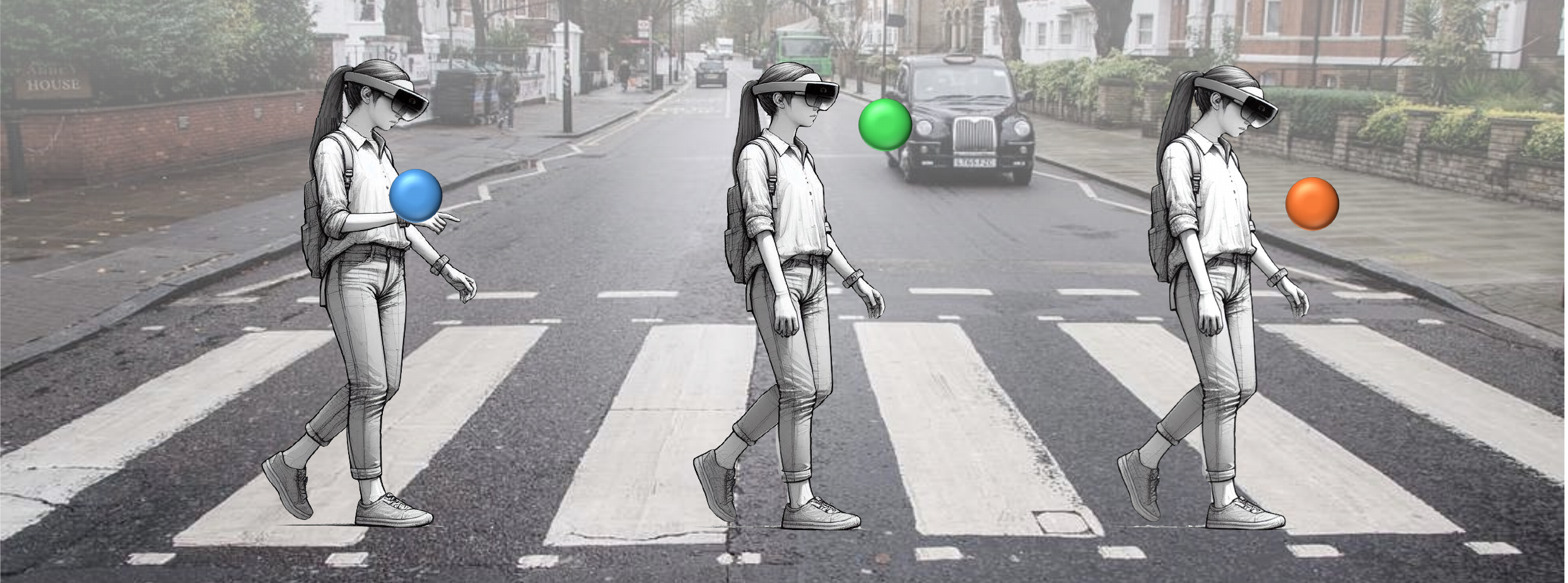}
	\caption{In this paper we explore the influence of different placements of AR content on walking and virtual task performance. We vary the placement of the content together with the virtual task difficulty and compare walking to stationary scenarios.}
	\Description{Teaser figure showing three people wearing augmented reality headsets while crossing a street. Each person is interacting with a different AR-anchored virtual object, represented by colored spheres (blue, green, and orange), symbolizing different AR anchoring methods: hand, head, and torso.}
	\label{fig:teaser}
\end{teaserfigure}

\maketitle

\section{Introduction}
\label{sec:introduction}







Attention is one of our most scarce and valuable resources, which we constantly need to shift and refocus. This poses a particular problem when we are not stationary in a familiar place but finding our way around the city \cite{farr_wayfinding_2012}, keeping an eye on traffic and avoiding obstacles on the pavement, while we are constantly inundated with a multitude of additional external information streams such as billboards and signs \cite{farbry_research_2004}. Over the last 15 years, the spread of smartphones has established another information channel that demands our attention while we are on the move \cite{liebherr_smartphones_2020}. Despite the inherent risks, people are using their smartphones even while walking \cite{beurskens_does_2013} to consume videos, read and answer text messages, or even play games \cite{stockel_cognitive_2020}. 
This can lead to dangerous walking situations, through inattentive walking, resulting in accidents in the past.
With the increasing suitability of \ac{AR} \acp{HMD} for everyday use, we will see a further surge in interaction in mobile scenarios and hence also when walking \cite{oulasvirta_interaction_2005}. Given the convergence of bits and atoms to a shared \ac{AR} and the corresponding disappearance of the need to look down and at an opaque device, \ac{AR} appears to be a better technical platform to support safe and enjoyable interaction while walking. However, the fundamental problems associated with interaction while walking also apply to \ac{AR} systems or are even exacerbated through increased use~\cite{lazaro_interaction_2021}. Frequent virtual interactions can lead to constant shifts of attention between the virtual and the physical world \cite{vortmann_eeg-based_2019, vortmann_exploration_2021}, affecting both the performance in the virtual task and the safety of walking \cite{chan_walking_2019}.

Previous work on the design of AR interfaces focused on user interactions in mostly static scenarios, examining aspects such as task performance and user engagement \cite{billinghurst_grand_2021}. %
Previous work also investigates the design and placement of \ac{AR} notifications in stationary scenarios \cite{lee_investigating_2022}, e.g., a social setting \cite{rzayev_effects_2020}, as well as in mobile scenarios \cite{kruijff_influence_2019, lee_exploring_2023} and how fast and accurate users can identify them.
Notifications are useful when users perform a primary task, and the system needs to redirect users' attention to a secondary task for a brief moment \cite{abdelnour_nocera_literature_2023}. Based on experience with smartphone systems, we assume that users will also interact with systems for longer periods of time while walking, leading to dual-task scenarios.
Here, related work only recently started investigating such dual-task scenarios for AR, e.g., by investigating windows placement for \ac{AR} video calls while on the go \cite{chang_exploring_2024}, working a virtual discrimination task while receiving navigation instructions on an AR HMD \cite{nenna_augmented_2021}, or how writing emails using an \ac{AR} \ac{HMD} while walking can be supported by \ac{AI} systems to help users focus more on the physical world \cite{zhou_glassmail_2024}. While these examples also use walking as a secondary task, they do not evaluate the impact on the walking behavior in the form of gait metrics like stride length, width, and frequency. Alongside path deviation, these, however, provide valuable insights into users' walking behavior, safety, and performance \cite{whittle_clinical_1996, danion_stride_2003}. 
Only a few examples also consider these measures, together with the effect of different anchoring in dual-task settings \cite{klose_text_2019}.
Consequently, a thorough understanding of the specific influences of AR content placement on attention management and physical and virtual task performance for dynamic mobile dual-task scenarios is lacking.


In this paper, we address the gap by systematically investigating the effects of varying the anchoring of AR content (hand, head, torso) and task difficulty on both walking performance (stride size, speed) and virtual task performance (reaction times, accuracy, miss rate) in a user study ($n=26$). During the study, participants engaged in a virtual working memory task with varying difficulty while following a dynamically changing path illuminated on the ground.
Our results can inform developers and designers to optimize mobile AR applications and highlight the interdependence of virtual and physical engagement and task performance in a mobile \ac{AR} scenario.
The findings of our work show a strong connection between physical and virtual task performance, which is in line with related work on shared attention \cite{al-yahya_cognitive_2011, plummer-damato_effects_2012, sarto_dual-tasking_2020}. Our findings further show the effect of anchoring of \ac{AR} content on both of the given tasks. Here, our results suggest that head-anchored content directly in the users \ac{FOV} reduces the users' task load and benefits their virtual and walking task performance most for comparable tasks. Hand anchoring, however, slowed down users in both tasks given and let them focus more deliberately on each.
In the broader context, this research enhances our understanding of user performance in mobile AR, guiding future developments in AR applications for everyday use, making them safer and user-friendly.

\section{Related Work}
\label{sec:relatedwork}
To highlight the relevance of our work, we provide an overview of interaction in mobile \ac{AR} environments, \ac{AR} content display, and dual-tasking in Human-Computer Interaction (HCI).

\subsection{Interaction on the Go}
With the advent of smartphones, interaction with information shifted from stationary systems, such as desktop PCs, into the physical world~\cite{lumsden_paradigm_2003}. The ubiquity of these devices led to widespread smartphone usage while walking~\cite{Yoshiki2017, marshall_mobile_2013, kwon_smartphone_2020}. However, focusing on a smartphone screen while walking reduces attention to physical surroundings~\cite{aagaard_mobile_2016, mourra_using_2020}, leading to collisions, obstacles, and dangerous situations for pedestrians and others~\cite{Schabrun2014a, Lin2017a, haolan_phone-related_2021}.

Voice-based interfaces can mitigate these risks by freeing the visual channel, enabling safer interaction while walking~\cite{peissner_can_2011}. Yet, they are hindered by noisy environments, limited privacy~\cite{Koelle2017}, and interference with interpersonal communication~\cite{Starner2002}, while also being unsuitable for certain digital content. For screen-based interaction, prior work has proposed solutions such as detecting texting while walking~\cite{Shikishima2018}, interrupting unsafe smartphone usage~\cite{Beuck2017}, or warning systems for hazardous situations~\cite{Hincapie-Ramos2013, Wang2017, Wen2015, Tang2016a}. Specialized approaches include support for texting~\cite{Kong2017} and video watching~\cite{Ahn2013}.

Beyond safety, walking induces situational impairments~\cite{sears03} that reduce interaction efficiency and accuracy~\cite{mourra_using_2020, xin_target_2023}, exacerbated by additional factors like carrying objects~\cite{Ng2013, Ng2015}. To address this, \citet{Kane2008} proposed enlarging buttons and text to preserve input performance, introducing \acp{WUI} to \enquote{compensate for the effects of walking on the usability of mobile devices}. This concept has since been extended to stabilize content~\cite{Rahmati2009}, adopt alternative keyboard layouts~\cite{Clawson2014}, and explore text input modalities like tilt~\cite{fitton_exploring_2013}.

While these solutions improve interaction safety and accuracy, they remain constrained by handheld devices, which require users to divert their gaze from the environment. To overcome these limitations, research has begun exploring \acp{HMD} for interaction while walking, as these devices allow users to maintain visual awareness of their surroundings~\cite{lucero_notifeye_2014}. For example, \citet{lages_walking_2019} studied adaptive interfaces for \acp{HMD} while walking, and \citet{chang_exploring_2024} investigated meeting interfaces for mobile users. Additional work has explored specific input techniques~\cite{Muller2020d} and use cases such as learning~\cite{ram_does_2022}.

Despite these advancements, it remains unclear how walking and interacting with digital information on \acp{HMD} influence each other, particularly given the unique ways \acp{HMD} present content.

\subsection{Display of AR Content}

Unlike traditional mobile screens like smartphones and smartwatches, AR HMDs display content directly in front of the user’s eyes, creating immersive experiences but also introducing visual distractions. Researchers have explored interface designs to enhance usability and task performance in various environments. For instance, \citet{luo_where_2022} and \citet{billinghurst_grand_2021} examined interface designs for managing attention and improving productivity in static settings, while other studies investigated displaying AR content on stationary and movable real-world objects to blend with physical reality \cite{han_blendmr_2023}.

In mobile \ac{AR}, notification design has been extensively studied, focusing on enabling users to process information efficiently during tasks such as walking or cycling \cite{plabst_exploring_2023}. \citet{kishishita_analysing_2014} and \citet{lee_investigating_2022} showed that well-placed notifications can help users maintain focus on primary tasks while receiving secondary information. However, these studies emphasize notification recognition efficiency rather than their impact on physical movement, cognitive load, or attention balance.

While prior work informed the design of AR interfaces for mobile environments, understanding how display placement affects physical tasks remains limited. Studies like \citet{klose_text_2019} explored AR content positioning in walking scenarios but relied on simple, static paths prone to learning effects and overlooked critical measures like walking stability and efficiency. Real-world dynamic environments, where users must adapt their movements while interacting with AR content, present greater challenges and remain underexplored.

Issues like visual clutter and perceptual fidelity add further complexity. Poorly designed notification systems can overwhelm users and degrade performance \cite{lazaro_interaction_2021, lee_effects_2020}. Placement and alignment, especially in mobile and social settings, significantly affect attention balance between virtual and physical environments \cite{rzayev_effects_2020}.

Understanding the effects of anchoring AR content to different body parts (hand, head, or torso) on both virtual and physical task performance, such as gait stability and efficiency, is critical. These interactions are key to improving the usability of AR in dynamic, everyday scenarios \cite{mack_head-anchored_2023}.

\subsection{Dual-Tasking in HCI}

Dual-tasking, the simultaneous performance of two tasks, is increasingly prevalent with the rise of mobile AR technologies, which require concurrent cognitive and motor engagement \cite{mckendrick_into_2016}. Examples include navigating complex environments with real-time AR navigation cues or technicians performing repairs using AR manuals. Understanding cognitive-motor interference is essential for designing user interfaces that balance cognitive load, ensure task performance, and maintain user safety.

Cognitive-motor interference occurs when simultaneous cognitive and motor tasks negatively impact each other’s performance \cite{al-yahya_cognitive_2011}. In AR, such interference can reduce efficiency and increase error rates in tasks requiring both mental and physical effort. For instance, \citet{prupetkaew_cognitive_2019} found that multitasking, such as texting while walking, significantly altered gait patterns and reduced walking speed, underscoring the importance of mobile interface designs that account for cognitive load during physical activities.

Visual attention is also markedly affected by dual-tasking. \citet{feld_visual_2019} observed that tasks like texting or verbal fluency while walking disrupted visual scanning behavior, reducing attention to the walking path and surroundings. This highlights the need for adaptive interfaces that mitigate cognitive-motor interference in dynamic environments.

Divided attention across the visual field is further challenged in complex environments. Increased attentional load impairs cognitive processing in the visual periphery \cite{spence_multisensory_2020, nenna_augmented_2021}. Walking speed also affects peripheral visual detection. \citet{cao_walking_2019} showed that walking enhances contrast sensitivity in the periphery, but this effect diminishes at higher speeds. Similarly, \citet{lim_dual_2015} found that nearly half of peripheral visual cues went undetected during walking and texting, with detection performance declining for stimuli farther from the central visual field.

\subsection{Research Gap}
Current research on AR interaction has largely focused on static and controlled environments, where the primary concern is how users engage with virtual content and how AR interfaces can allow for efficient performance. In addition, related work examined AR notification systems and related dual-task scenarios, such as walking while receiving virtual notifications. However, these studies often overlook key aspects of physical performance, particularly how AR content anchoring affects walking behavior and influences cognitive-motor interference in dynamic settings.

Most research relies on simplified or static walking paths, which fail to capture the complexity of real-world environments. Consequently, little is known about the effects of AR content anchoring on both physical and virtual task performance in more unpredictable, dynamic conditions. The interaction between physical navigation and virtual engagement remains under-explored, as does the impact of AR content anchoring on walking efficiency and stability. While static world-anchoring and object-based anchoring have been explored in prior work, these approaches often lack adaptability in dynamic and unpredictable contexts. 

To address this gap, we investigate how different AR content anchoring positions around the users' bodies affect their performance in both virtual tasks and physical walking in dynamic environments. Specifically, we explore how head-, torso-, and hand-anchored AR content influences virtual task efficiency, walking performance, and cognitive load under varying task difficulties and mobility conditions. By focusing on body-centric anchoring, we provide a clear understanding of its impact on user performance in dynamic environments, offering a basis for future studies to explore adaptive anchoring strategies. Our study introduces a dynamic walking task that mirrors the unpredictability of real-world conditions, allowing us to examine how AR content placement interacts with physical movement.

\section{Methodology}
\label{sec:methodology}

\begin{figure}[t]
    \centering
    \includegraphics[width=1.0\linewidth]{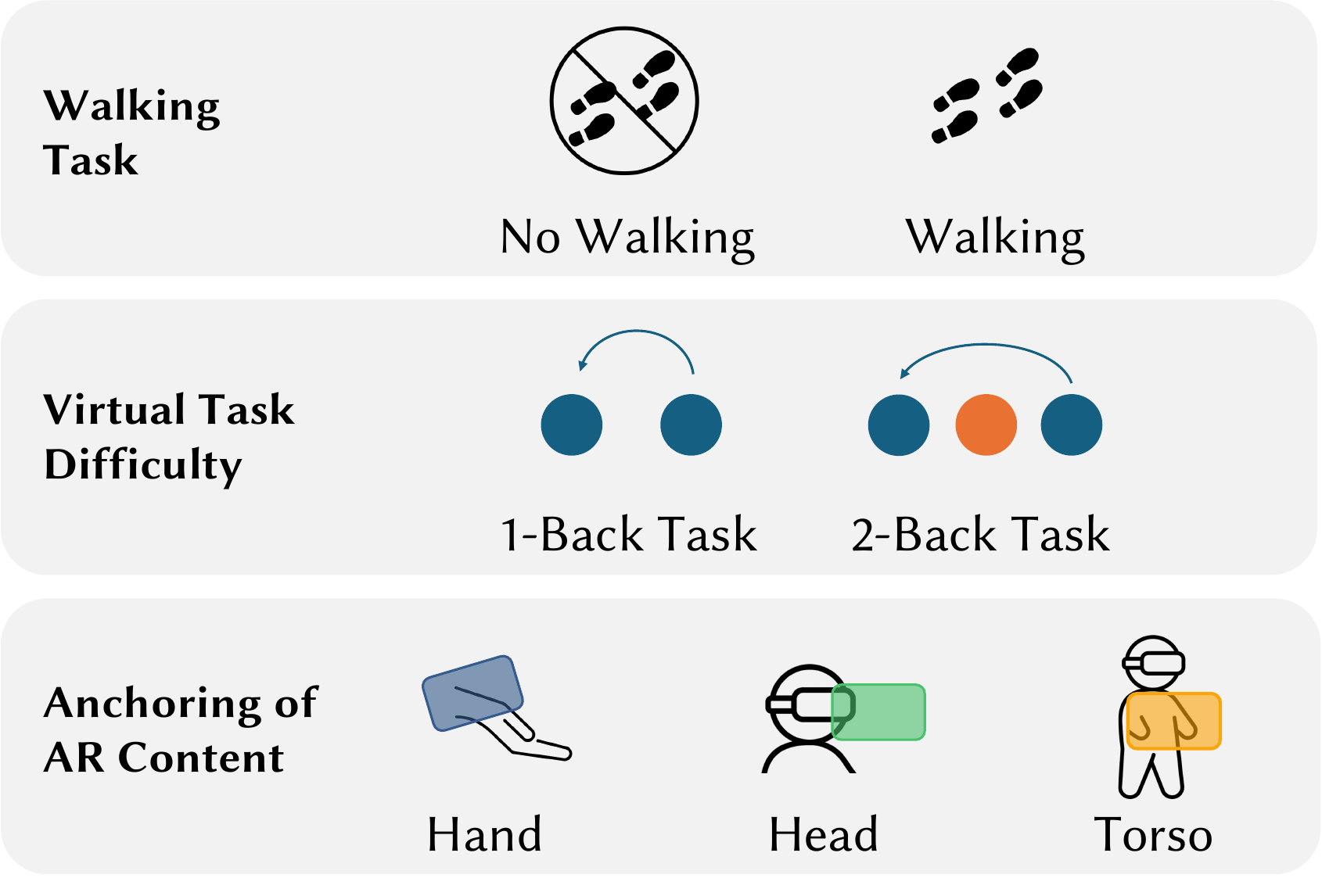}
    \caption{The three independent variables of our study \ivPhysTask{} (\NoWalking{} / \Walking{}), \ivDifficulty{} (\nback{1} / \nback{2}), and AR Content \ivAnchor{} (\pHand{} / \pHead{} / \pTorso{}). The \NoWalking{} conditions act as baseline conditions to assess participants' single task virtual task performance.  An additional \conWalkingonly{} condition acts as a baseline condition to assess participants' single task physical task performance.}
    
    \label{fig:IV_Figure}
    \Description{The figure shows a schematic representation of the experimental design, divided into three sections: Walking Task, Virtual Task Difficulty, and Anchoring of AR Content. The Walking Task is illustrated with two icons: "No Walking," represented by a pair of stationary footprints, and "Walking," shown with a pair of moving footprints. Virtual Task Difficulty is represented with two icons: the "1-Back Task" (a simple task) with two dots connected by a curved arrow, and the "2-Back Task" (a more complex task) with three dots connected by a curved arrow. Anchoring of AR Content is represented by three icons: "Hand" anchoring, with an image of a hand holding a device; "Head" anchoring, with an image of a head wearing AR glasses; and "Torso" anchoring, with a figure wearing an AR display attached to their torso.
    }
\end{figure}
To study the influence of AR content placement in a mobile scenario regarding its effect on virtual and walking task performance and user experience, as well as their interplay, we formulate two research questions and evaluate them in our user study. We ground our RQs in the related work presented in \autoref{sec:relatedwork} and formulate them as follows.

\begin{itemize}
    \item [\textbf{RQ1:}] How does AR content placement affect users' virtual task performance and demand for stationary and walking\,scenarios?
    
    \item [\textbf{RQ2:}] How does AR content placement affect users' walking performance and demand for varying virtual task difficulties?
\end{itemize}

\subsection{Study Design}
We used a within-participants experimental design. We vary three independent variables \ivPhysTask{} (two levels : \NoWalking{} / \Walking{}), \ivDifficulty{} (two levels : \nback{1} / \nback{2}), and AR Content \ivAnchor{} (three levels : \pHand{} / \pHead{} / \pTorso{}).
Overall, our study consists of $2 \times 2 \times 3$ conditions plus a \conWalkingonly{} baseline condition, resulting in 13 conditions. To avoid learning and carry-over effects, we counterbalance the order of conditions using a Balanced Latin Square Design \cite{wang_construction_2009}.

To allow a focus on the placement of the content without interfering with suitable input techniques, we gave participants a physical remote presenter with 2 input buttons required to solve the virtual \nback{n} task.

\subsubsection{Independent Variables}

During our experiment, we manipulate three independent variables.

\paragraph{Task Setting}
The first is the \ivPhysTask{}, which has two levels: \NoWalking{} and \Walking{}. The \NoWalking{} condition serves as a baseline to assess participants' virtual task performance in a single-task setting, while in the \Walking{} condition, we vary the path dynamically to prevent participants from memorizing a specific route, a mechanism that we explain later in the paper.

\paragraph{Virtual Task Difficulty}
The second variable is the \ivDifficulty{}  of the virtual \nback{n} task, which includes two levels: \nback{1} and \nback{2}.  In this established working memory task \cite{chiossi_adapting_2023, chiossi_virtual_2022}, participants have to compare the current item to the item n items before. We also use a \conWalkingonly{} condition without a virtual task as a baseline for assessing the users' single-task walking performance.

\paragraph{AR Anchoring}
The last independent variable is the \ivAnchor{} of the virtual content, which has three levels: \pHand{}, \pHead{}, and \pTorso{}. In the \pHand{} condition, the content is anchored to the participant’s left hand, moving naturally with their hand movements. The \pHead{} condition fixes the content within the participants' FOV, ensuring it always stays in front of them without requiring head movement to follow it. In the \pTorso{} condition, the content is attached to the participant’s chest, moving with their torso as they walk. Our study, therefore, covers the most typical types of mobile AR interfaces that follow the user~\cite{klose_text_2019, chang_exploring_2024}. The selected three anchor points allow us to examine how different body-based references affect both cognitive and motor task performance. 

\subsubsection{Dependent Variables}
\label{sec:methodology:dvs}

During each condition, we collect data on both walking and virtual task performance, as well as subjective measures. 

\paragraph{Walking Task Performance}
For the walking task performance, we evaluate participants' walking accuracy and efficiency through several metrics. First, we use \dvWalkingError{} as a measure of walking accuracy, where higher error rates indicate less precise navigation along the highlighted path. We calculated the \dvWalkingError{} as the distance of the foot to the highlighted target walking path to measure the participants' walking accuracy.
Additionally, we assess walking efficiency through \dvStrideDuration{}, \dvStrideLength{}, and \dvStrideWidth{}. We calculate the \dvStrideDuration{} as the time of a step (when one foot is in the air). Here, longer \dvStrideDuration{} reflects slower steps.

\begin{figure}[h]
    \centering
    \includegraphics[width=0.7\columnwidth]{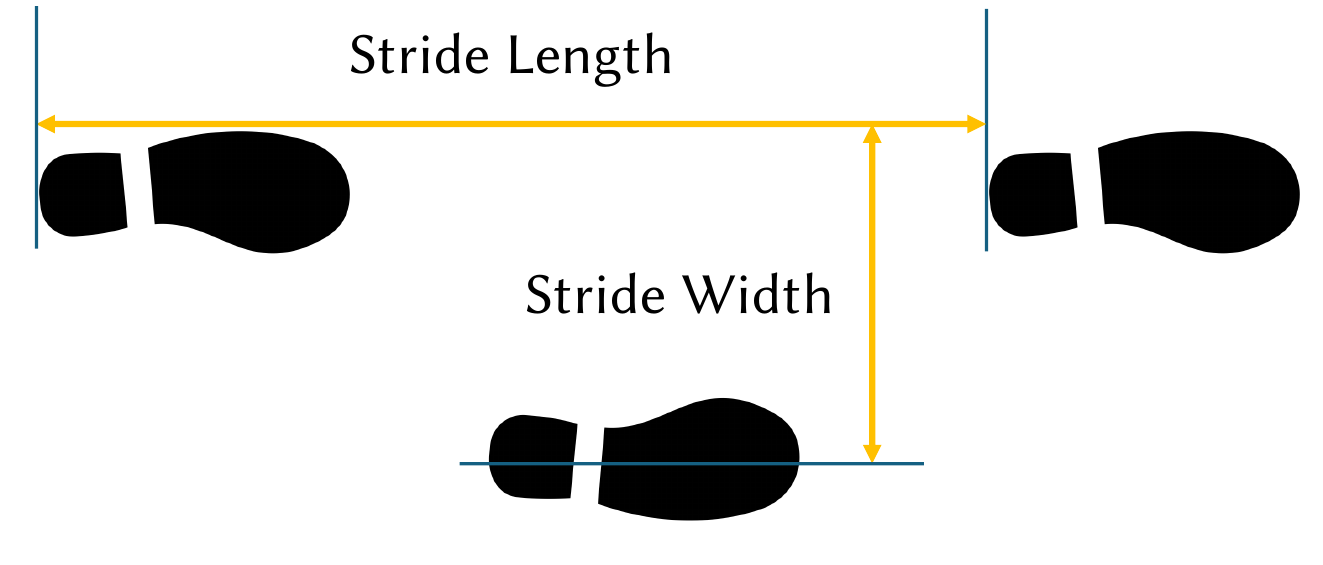}
    \caption{Explanation of the Stride Length and Stride Width. Stride Lenght was computed as the distance between two successive steps of the same foot, while stride width was calculated as the perpendicular distance between the feet during consecutive steps. These measures were recorded to assess changes in walking performance under varying task conditions.}
    \Description{A figure showing the calculation of stride length and stride width. Stride length is represented as the distance between two consecutive steps of the same foot, while stride width is shown as the perpendicular distance between the left and right feet during walking. This visual illustrates how walking performance measures were computed during the analysis of the study.}
    \label{fig:stride-length-width}
\end{figure}

We calculate the \dvStrideLength{} as the distance between the two feet on the ground in the walking direction (see \autoref{fig:stride-length-width}). Here, a greater \dvStrideLength{} indicates larger steps.
Furthermore, we calculate the \dvStrideWidth{} as the distance between the two feet on the ground orthogonal to the walking direction (see \autoref{fig:stride-length-width}), with wider steps suggesting an increased need for walking stability\,\cite{bauby_active_2000, dean_effect_2007}.

\begin{figure*}[t!]
    \vspace{-1em}
	\begin{minipage}[t]{.4\linewidth}
		\centering
        \includegraphics[width=\linewidth]{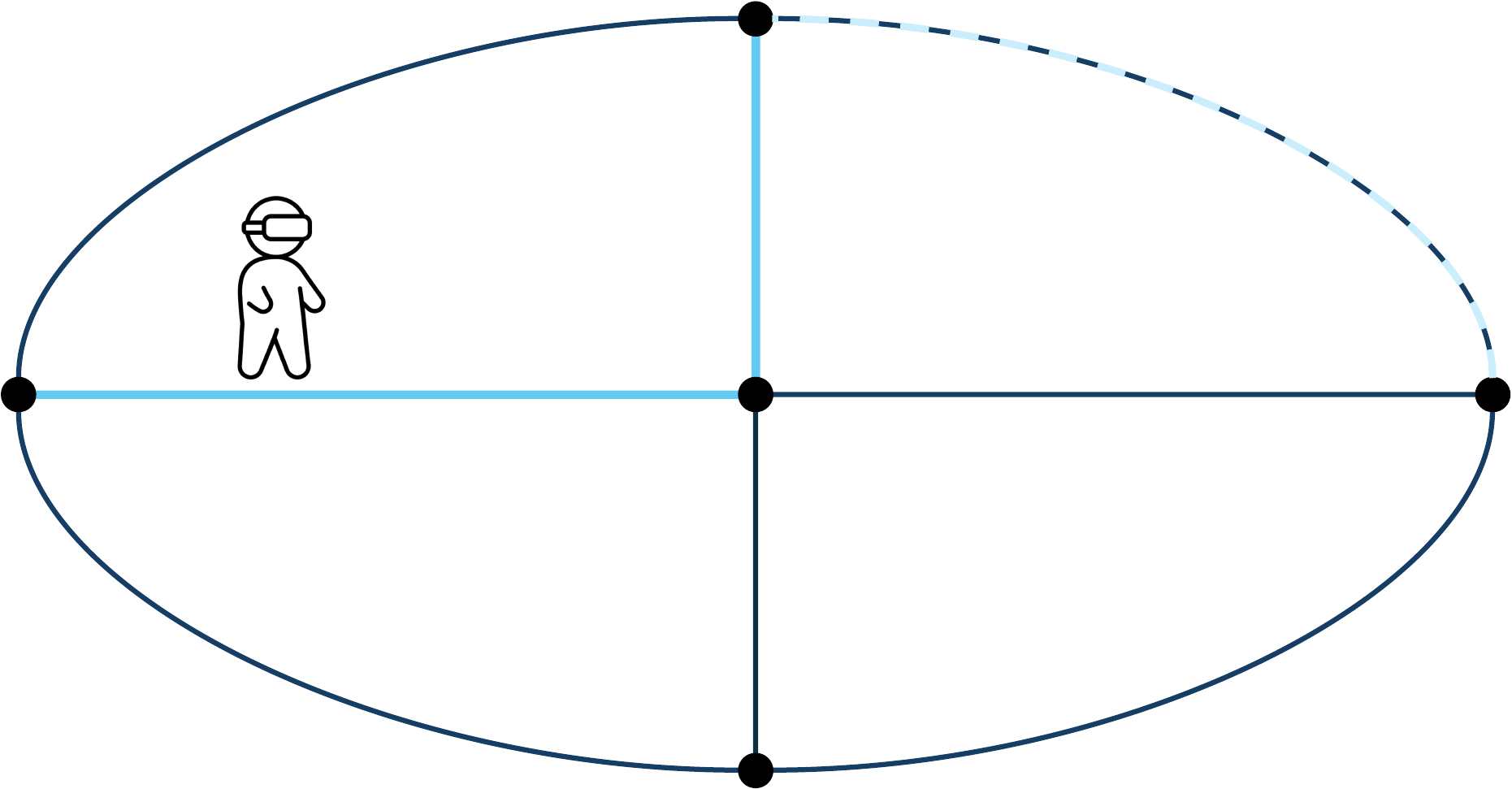}
		\subcaption{}
        \label{fig:walkingpath1}
	\end{minipage}%
\hspace{1cm}
        \begin{minipage}[t]{.4\linewidth}
		\centering
        \includegraphics[width=\linewidth]{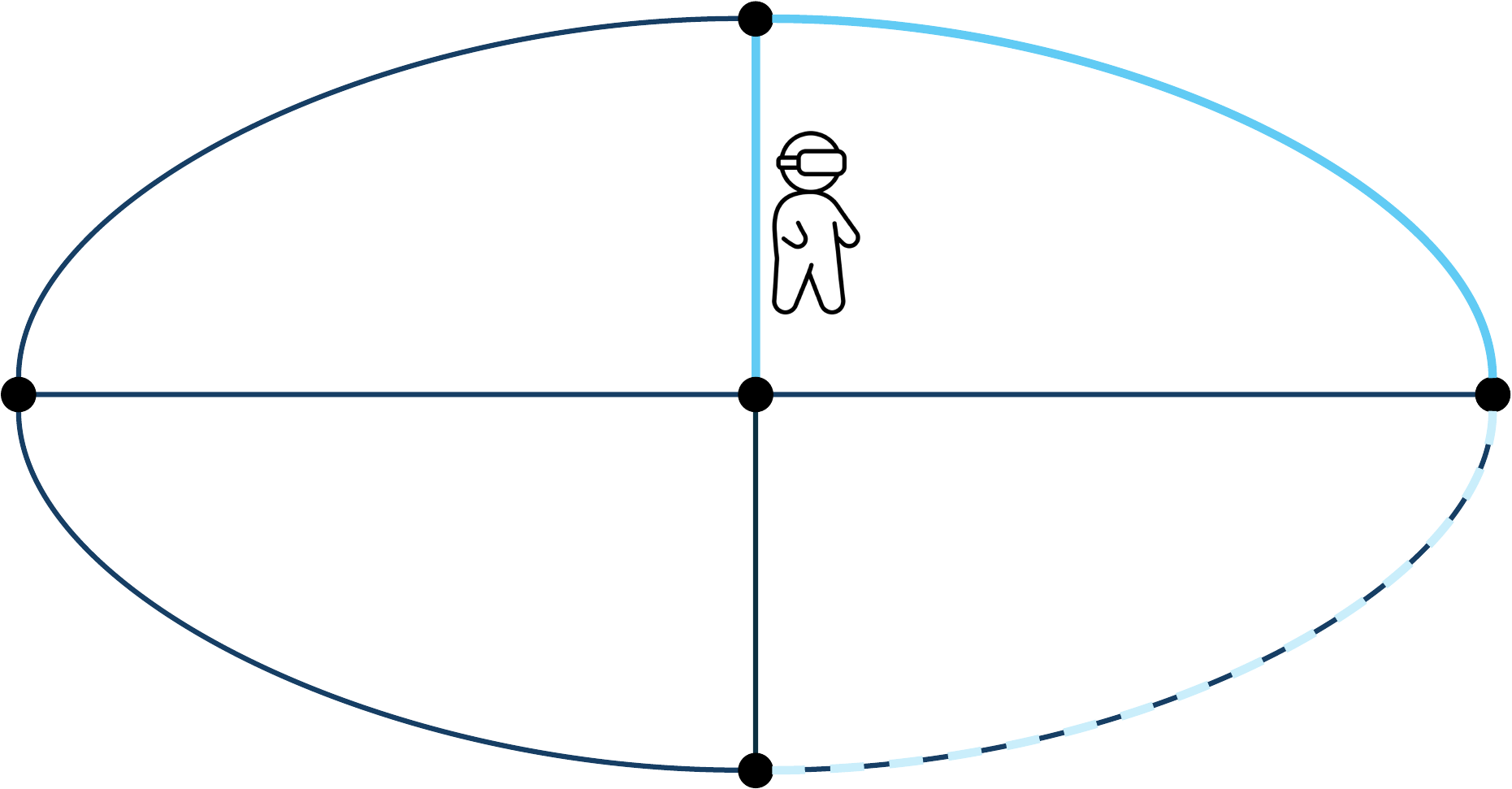}
  		\subcaption{}
        \label{fig:walkingpath2}
	\end{minipage}%
	\caption{A top view schematic of the walking area. The width is 6.8m and the height 3.4m. The light blue elements show the path generation during the experiment, the dotted light blue line shows the respective next element to illuminate. (a) The participant starts at the first node and sees the first two path sections. (b) Once the participant reaches the second node, the next path section lights up and the last section turns off. This continuous to ensure that before reaching the node, the path section behind the next node is already illuminated.}
	\Description{A top view schematic of the walking area used in the experiment. The width of the area is 6.8 meters, and the height is 3.4 meters. The light blue elements show the path generation during the experiment, while the dotted light blue line represents the illuminated path sections. In diagram (a), the participant starts at the first node and can see the first two path sections. In diagram (b), after reaching the second node, the next path section lights up, and the last section turns off. This sequence continues to ensure that the path section behind the next node is illuminated before reaching it.}
	\label{fig:walkingpath12}
\end{figure*}

\paragraph{Virtual Task Performance}
For virtual task performance, we measure participants' accuracy and response times in the n-back task. We calculate the \dvAccuracy{} by dividing the number of correctly answered n-back items by the total items answered in each condition. In addition, we capture the \dvAnswerTime{} in seconds it took participants to respond to each n-back item, with items answered in more than 4 seconds being marked as missed. The \dvMissedAnswerRate{} reflects the proportion of missed answers, i.e., users did not provide an answer within 4 seconds, compared to the total number of responses.

\paragraph{Subjective Measures}

Finally, we gather subjective measures to assess participants' perceived task demand and focus. We use the \ac{RTLX} as a measure of overall task demand \cite{hart_nasa-task_2006}. Additionally, we administer a custom 7-point Likert questionnaire (strongly disagree to strongly agree) with questions tailored to evaluate participants’ walking and virtual task performance, as well as their perceived demand and focus during the experiment consisting of the following 6 items: "My Performance on the Virtual Task Was Very Successful", "I Found It Hard to Focus on the Virtual Task", "My Performance on the Physical Task Was Very Successful", "I Found It Hard to Focus on the Physical Task", "The Physical Task Was More Demanding Than the Virtual Task", and "The Virtual Task Distracted Me From the Physical Task".

\subsection{Apparatus}

The apparatus for this study comprised the Microsoft HoloLens 2 to render the virtual task, a tower PC for data recording (Intel Core i7 with 3.00GHz, 32GB RAM), and a dynamic light path system. We developed the virtual task with the Unity game engine (Version 2021.3.34f1). For recording users' behavioral performance in the virtual n-back task, we employed a Bluetooth Presenter \footnote{\url{https://news.microsoft.com/wp-content/uploads/prod/sites/646/2022/10/Microsoft-Presenter-Fact-Sheet.pdf}} connected to the recording PC. We let participants choose their preferred hand for the interaction with the remote presenter. We implement a setup to generate a dynamic path during the experiment, ensuring participants need to pay attention to their physical environment and react to the new path elements appearing. We also include a VIVE tracking system to measure the walking performance of the users and use the position as input for path generation.

\subsubsection{Motion Tracking}
Participants’ movements were recorded using the VIVE tracking system together with three Tundra Trackers (90Hz)\footnote{\url{https://tundra-labs.com}}, positioned on each foot and the sternum. Four VIVE base stations, placed around the walking space, provided optical infrared laser-based distance measurements through trigonometry \cite{bauer_accuracy_2021}. The position and orientation of the trackers and HMD were calibrated at the start of each session to ensure accurate tracking. To do this, we placed the HMD and trackers at a predefined, central point in the room before starting the calibration step, hereby aligning the position and orientation for both coordinate systems. Previous studies validated the tracking accuracy of the VIVE Tracking system with tracking errors in the submillimeter range for comparable dynamic experiments\,\cite{kuhlmann_de_canaviri_static_2023} and sample-to-sample jitter in the submillimeter range as well \cite{niehorster_accuracy_2017}. 

\subsubsection{Dynamic Walking Path}
To create a dynamic, real-world path, we used Electroluminescence (EL) wire\footnote{\url{https://en.wikipedia.org/wiki/Electroluminescent_wire}}, chosen for its durability and ability to be walked on without damage. The bright blue EL wire was visible in standard room lighting and through the HMD visor. We controlled the lights using an Arduino UNO\footnote{\url{https://docs.arduino.cc/hardware/uno-rev3/}} microcontroller and a transistor-based power supply. The system dynamically updated the path using live tracking data from the torso, which was sent from Unity to the microcontroller. As participants reached an intersection, one of the two unlit path segments was randomly activated, creating an unpredictable and constantly changing path.

\subsubsection{Anchoring Reference Points}

\paragraph{Hand Anchoring}
For the anchoring of virtual content, we used specific real-world reference points on the user's body. In the \pHand{} condition, we anchored the content to the left hand by tracking it using the Hololens 2 SDK, allowing the virtual elements to follow the participant’s hand movements. Because of the small form factor of the remote presenter, participants could also hold the presenter in their preferred hand without interfering with the anchoring. The distance and size of the displayed content changed dynamically based on the user's hand position. The hand, according to the Human Engineering Design Data Digest \cite{poston_human_2000}, is between 60–90 cm from the head of the user. This allowed us to stay within the 80 cm recommended distance from the head by \citet{hussain_effects_2023}, i.e., within users' field of view (approximately 43$^\circ$ horizontal and 29$^\circ$ vertical as typical for AR headsets like the HoloLens 2). The virtual sphere, representing the anchored content, was instantiated with dimensions of .1 x .1 x .1 units in Unity's measurement system, equivalent to a sphere with a 10\,cm diameter. It was displayed directly above the left hand, positioned .1 units (10 cm) above the hand in the local y-axis. When viewed at a distance of 80 cm, the sphere occupied approximately 16\% of the vertical FOV and 24\% of the horizontal FOV of the HoloLens 2. This placement aligns with ergonomic recommendations from the Microsoft Mixed Reality Design Guidelines\footnote{\url{https://learn.microsoft.com/en-us/windows/mixed-reality/design/hand-menu##hand-menu-placement-best-practices}}, which suggests placing menus above the hand to reduce the need for raising the arm excessively, thereby minimizing user fatigue during prolonged tasks.

\paragraph{Head Anchoring}
For the \pHead{} condition, we displayed the virtual sphere with a consistent diameter of 10\,cm (scale .1) at a distance of approximately 100\,cm in front of the head in the direction of view from the HoloLens 2 head tracking system. This placement ensured that the content remained within the center of the participant's FOV, as calculated by the Hololens 2 SDK while maintaining consistent visibility and alignment with ergonomic guidelines for head-anchored content.

\paragraph{Torso Anchoring}
In the \pTorso{} condition, we anchored the sphere representing the anchored virtual content to a Tundra tracker attached to the participant's chest. We carefully positioned the chest tracker on the sternum to accurately reflect torso movements\,\cite{westlund_motion_2015}. Again, we position the virtual sphere with a 10\,cm diameter 100\,cm in front of the reference point at the chest. The tracker provided the positional data necessary to anchor the virtual content in alignment with the participant’s torso, thus rendering the virtual content move naturally with their body as they navigated the dynamic path. This placement ensured that the virtual content remained aligned with the participant’s natural line of sight, 15$^\circ$–20$^\circ$ below the normal horizontal line of sight during forward locomotion, as recommended by \citet{moore_human_2001}. This height placement was chosen to minimize the need for excessive head or gaze movement while navigating, reducing cognitive load and ergonomic strain, and maintaining visibility of the dynamic path.

\subsection{Tasks}
In our study, we tasked participants to complete a virtual and a walking task simultaneously, creating a dual-task scenario designed to test the cognitive-motor interference under different AR anchoring conditions. Each condition lasted 2 minutes.

\subsubsection{Virtual Task on AR Headset}
For the virtual n-back task\,\cite{kirchner_age_1958, jaeggi_does_2003} used in our study, we adapted it from an established n-back task from related work\,\cite{chiossi_virtual_2022, chiossi_adapting_2023}, originally designed for a VR environment. We presented participants with spheres of four different colors: green \textcolor[rgb]{0,0.5,0}{\texttt{\#008000}}, red \textcolor{red}{\texttt{\#BF1818}}, blue \textcolor{blue}{\texttt{\#0000FF}}, and black \textcolor[rgb]{0,0,0}{\texttt{\#000000}}, following the recommendations by\,\citet{mcmillan_self-paced_2007}.
We generated the color sequence randomly, and participants interacted with these spheres using a presenter as input. 

The task required participants to push a button upon sphere presentation and decide whether it matched the color of the sphere presented one (1-back task) or two (2-back task) steps earlier. If the sphere matched the color, participants had to press the right button on the presenter; otherwise, they had to press the left button. In line with the task setup in related work \cite{chiossi_virtual_2022, chiossi_adapting_2023}, participants had up to 4 seconds to react and categorize the sphere. If the sphere was not categorized within this time frame, we classified it as a miss, the sphere would disappear, and the next sphere would automatically appear. To provide continuous feedback, participants received accuracy updates every 20 spheres and were instructed to maintain at least 90\% accuracy throughout the task. We recorded missed spheres or wrong button presses as errors.

\subsubsection{Walking Task}
In the walking task, we instructed participants to accurately follow a dynamically changing path in real-time. We designed this dynamic path to keep participants focused on both physical movement and the evolving environment. The path was represented by an illuminated EL wire, which was visible through the HMD visor and was durable enough for participants to step on without interference.
As participants walked, the path ahead dynamically updated. At each intersection, a random new segment of the path illuminated once participants reached the previous intersection, forcing them to pay close attention and make real-time adjustments to their route. The task was unpredictable, requiring constant focus on the next available path. This dynamic setup, controlled by live tracking data from the VIVE system, ensured that participants could not memorize the path and had to react to new path elements as they progressed, simulating a demanding dual-task scenario where both physical navigation and attention were required.

\begin{figure*}[t!]
	\includegraphics[width=\linewidth]{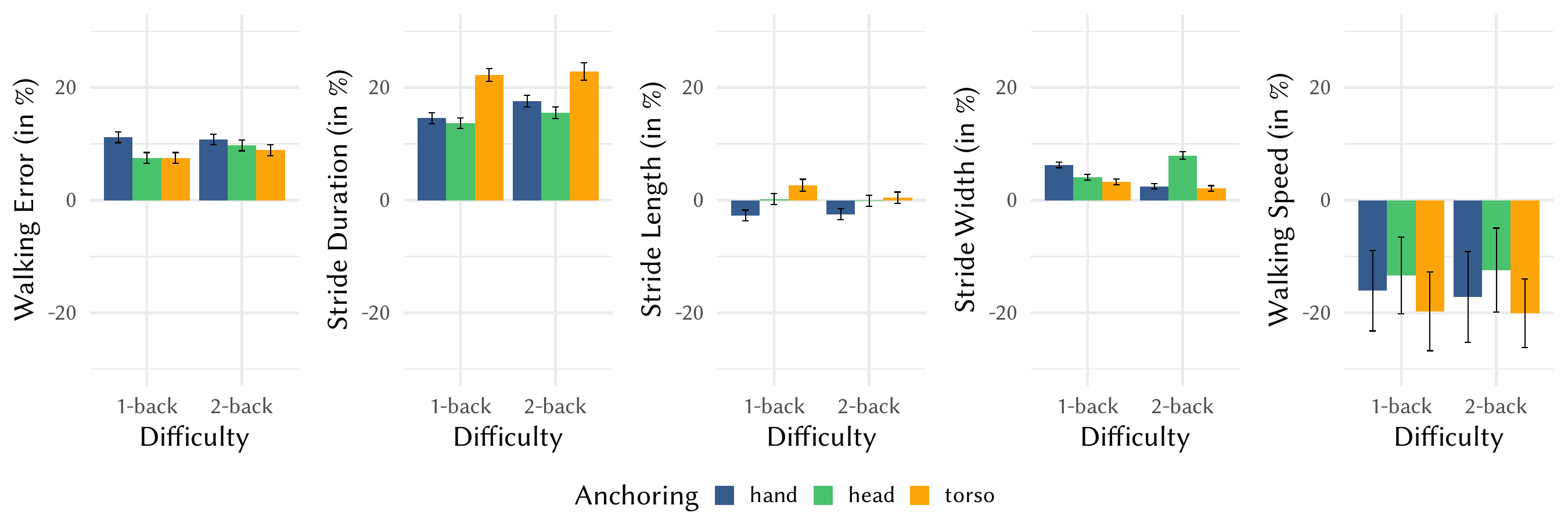}

	\begin{minipage}[t]{.2\linewidth}
		\centering  
        \vspace{-1em}
		\subcaption{\dvWalkingError{}}\label{fig:results:walking:walking_error}
	\end{minipage}%
    \begin{minipage}[t]{.2\linewidth}
		\centering
        \vspace{-1em}
		\subcaption{\dvStrideDuration{}}\label{fig:results:walking:stride_duration}
	\end{minipage}%
     \begin{minipage}[t]{.2\linewidth}
		\centering
        \vspace{-1em}
		\subcaption{\dvStrideLength{}}\label{fig:results:walking:stride_length}
	\end{minipage}%
	\begin{minipage}[t]{.2\linewidth}
		\centering
        \vspace{-1em}
		\subcaption{\dvStrideWidth{}}\label{fig:results:walking:stride_width}
	\end{minipage}%
	\begin{minipage}[t]{.2\linewidth}
		\centering
        \vspace{-1em}
		\subcaption{\dvWalkingSpeed{}}\label{fig:results:walking:speed}
	\end{minipage}%
	\caption{The dependent variables for the walking performance. All measurements are normalized using the baseline walking condition of the respective participant and, thus, represent the difference to the normal walking in percent. All error bars depict the standard error. (a) The walking error slightly increased for hand anchoring across all n-back levels. (b) Stride duration was significantly longer for torso anchoring, particularly with the 2-back task. (c) Stride length was shorter for hand anchoring and increased for head and torso anchoring, suggesting hand anchoring induces more cautious walking patterns. (d) Stride width increased with head anchoring under more difficult tasks, reflecting a potential need for greater stability when attention is focused on the virtual task. (e) Walking speed did not change significantly between the conditions.}
	\label{fig:results:walking}
 \Description{The figure consists of five bar charts illustrating the impact of AR content anchoring (hand, head, torso) and task difficulty (1-back, 2-back) on various walking performance metrics. Chart (a) "Walking Error" shows slight increases in error for hand anchoring compared to head and torso across both task difficulties. Chart (b) "Stride Duration" shows that stride duration increases with task difficulty, with torso anchoring showing the longest durations. Chart (c) "Stride Length" shows that hand anchoring leads to shorter strides, while head and torso anchoring result in longer strides. Chart (d) "Stride Width" shows wider strides with head anchoring, particularly in more difficult tasks, indicating a need for greater stability. Chart (e) Walking speed did not change significantly between the conditions.}
\end{figure*}

\begin{table*}
\centering
\caption{Mean Values (M) and Standard Deviation (SD) of the walking data analysis. All values are relative to the baseline and in\,\%.}
\Description{A table providing an overview of the Mean Values and Standard Deviation of the walking data analysis.}
\resizebox{\ifdim\width>\linewidth\linewidth\else\width\fi}{!}{
\begin{tabular}{llrrrrrrrrrr}
\toprule
\multicolumn{1}{c}{} & \multicolumn{1}{c}{} & \multicolumn{2}{c}{Walking Error} & \multicolumn{2}{c}{Stride Duration} & \multicolumn{2}{c}{Stride Length} & \multicolumn{2}{c}{Stride Width} & \multicolumn{2}{c}{Walking Speed} \\
\cmidrule(l{3pt}r{3pt}){3-4} \cmidrule(l{3pt}r{3pt}){5-6} \cmidrule(l{3pt}r{3pt}){7-8} \cmidrule(l{3pt}r{3pt}){9-10} \cmidrule(l{3pt}r{3pt}){11-12}
Difficulty & Anchoring & M & SD & M & SD & M & SD & M & SD & M & SD\\
\midrule
1-back & Hand & 111.18 & 71.47 & 114.58 & 73.91 & 97.28 & 73.10 & 106.27 & 37.33 & 83.93 & 35.04\\
1-back & Head & 107.52 & 71.41 & 113.66 & 69.97 & 100.21 & 76.23 & 104.09 & 38.98 & 86.61 & 33.44\\
1-back & Torso & 107.49 & 70.50 & 122.27 & 82.13 & 102.67 & 75.84 & 103.25 & 38.55 & 80.28 & 34.33\\
2-back & Hand & 110.78 & 70.72 & 117.61 & 76.57 & 97.52 & 72.27 & 102.48 & 36.74 & 82.81 & 39.39\\
2-back & Head & 109.74 & 72.22 & 115.52 & 76.78 & 99.92 & 74.19 & 107.93 & 48.99 & 87.60 & 36.45\\
2-back & Torso & 108.91 & 72.08 & 122.86 & 114.76 & 100.44 & 73.58 & 102.10 & 35.10 & 79.93 & 30.00\\
\bottomrule
\end{tabular}}
\end{table*}


\subsection{Participants}
26 participants (16 female, 10 male, none diverse) voluntarily participated in the study. The mean age was 27.1 years, ranging from 20 to 58 years. 
All participants could walk without aid, had no orthopedic, neuromuscular, or dementia disorders, and were independent in daily activities. They reported normal or corrected-to-normal vision. We conducted this study in line with our institution's ethics guidelines. All participants provided written informed consent before starting the study and received a compensation of 15€ for their participation.

\subsection{Procedure}
After welcoming our participants, they filled out an informed consent form for their participation in our study and filled out a demographics survey. We then explained the task and devices used. 
For the \nback{n}, they were specifically instructed to balance accuracy and speed together with the walking task, rather than prioritizing one over the other, to achieve an optimal performance trade-off based on \citet{rival_effects_2003}. This approach was designed to encourage participants to focus on both the quality and efficiency of their responses, ensuring a balanced performance. Participants then put on the \ac{AR} \ac{HMD} as well as the trackers for feet and torso. We assisted participants if required here. Before starting with the task, we ensured the correct positioning and fit of the equipment and resolved potential questions of the participants. 
After this, participants started the experimental phase consisting of thirteen counterbalanced experimental blocks, each with a task time of 2 minutes. In between conditions, participants took off the \ac{HMD} and filled out the NASA TLX and custom Likert Scales before continuing with the next condition.
Overall, the experiment lasted one hour and thirty minutes.

\section{Results}
\label{sec:results}

\begin{figure*}[t!]

	\includegraphics[width=\linewidth]{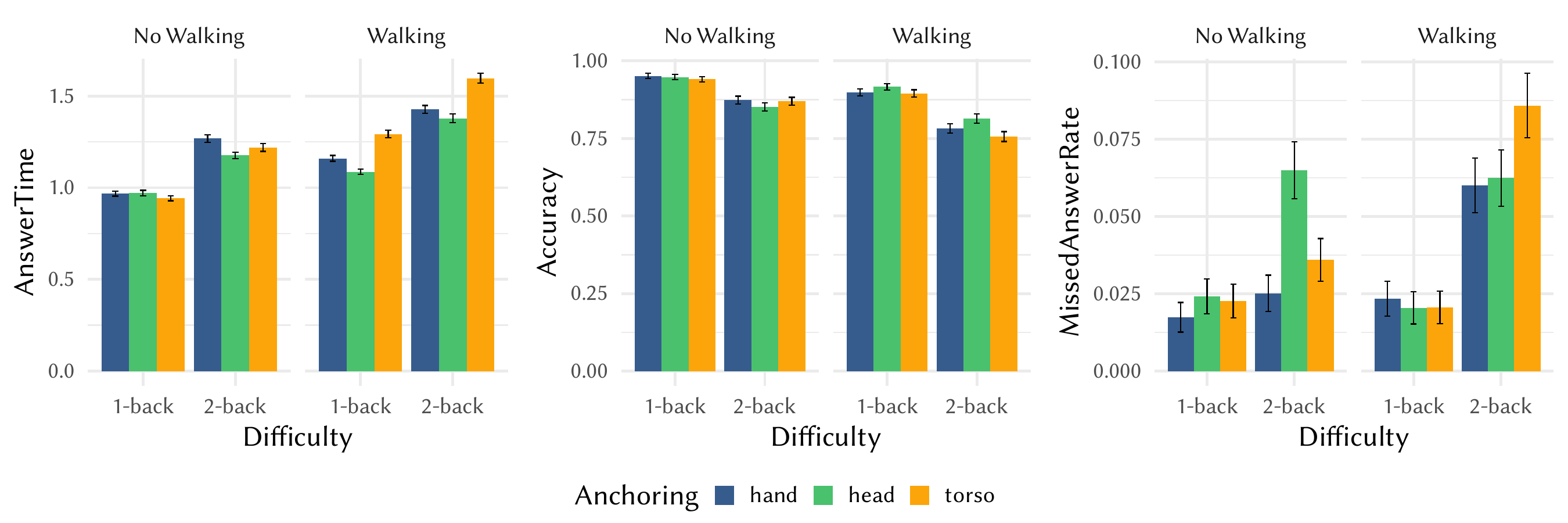}
	\begin{minipage}[t]{.3\linewidth}
		\centering
        \vspace{-1em}
		\subcaption{\dvAnswerTime{}}
        \label{fig:results:AnswerTime}
	\end{minipage}%
    \begin{minipage}[t]{.3\linewidth}
		\centering
        \vspace{-1em}
	    \subcaption{\dvAccuracy{}}
        \label{fig:results:CorrectAnswerRate}
	\end{minipage}%
    \begin{minipage}[t]{.3\linewidth}
	   \centering
       \vspace{-1em}
	   \subcaption{\dvMissedAnswerRate{}}
        \label{fig:results:MissedAnswerRate}
	\end{minipage}%
	\caption{The mean results for (a) \dvAnswerTime{} and (b) \dvAccuracy{} and (c) \dvMissedAnswerRate{} as a bar chart plot. The error bars indicate the standard error. Answer times increased when the n-back level was higher, particularly in walking conditions, and were slowest for torso anchoring. Accuracy declined slightly with task difficulty and when walking, but there was no significant effect from anchoring. Missed answer rates increased substantially with higher task difficulty, especially during walking, with hand anchoring resulting in fewer missed answers compared to head and torso anchoring in more difficult tasks.}
	\Description{The figure presents three bar charts illustrating the impact of AR content anchoring (hand, head, torso) and task difficulty (1-back, 2-back) on virtual task performance, with and without walking. The first chart shows "Answer Time," where response times increase with higher task difficulty and walking, and are slowest for torso anchoring. The second chart shows "Accuracy," where performance slightly decreases with task difficulty and walking, but anchoring has no significant effect. The third chart shows "Missed Answer Rate," which increases with higher task difficulty and walking, with fewer missed answers for hand anchoring compared to head and torso anchoring in the most challenging conditions.}
	\label{fig:plot_answertime_rate}
\end{figure*}

In this section, we present the results of our user study. We start with the results of the walking performance, followed by the virtual task performance, and lastly, with the subjective measures. Since each data type requires a different analysis, we explain the used analysis before presenting the respective results.

\subsection{Walking Performance}
To explore the influence of simultaneous digital interaction during walking on the gait, we analyzed various measures to quantify changes in the participants' gait as outlined in section \ref{sec:methodology:dvs}. For this analysis, we only considered the conditions that included walking while interacting with the n-back task. 

To collect these dependent variables, we started with the logged raw movement data for both feet. Based on this data, we identified the data points that represent the individual steps of each foot. To do this, we screened for regions in the time sequence of this data at which the height of the foot (as the position of the tracker on the foot along the height axis) reached both a local minimum and remained roughly stable over a period of more than 0.1 seconds. Using this process, we identified a total of 47145 footsteps. Based on the identified footsteps, we calculated the dependent variables analyzed below. We normalized all dependent variables to the calibration walk of the respective participants without interaction with a digital task to account for personal differences. Therefore, all measurements indicate percentages relative to the participants' normal gait. We removed the data of 3 participants due to technical problems that prevented us from reliably tracking their movement data. Further, we removed 2747 of the remaining 43904 data points we identified as outliers (outside $1.5xIQR$ below the first or above the third quartile).

For all dependent variables, we computed linear mixed-effects models (LMEs) with \ivDifficulty{}, \ivAnchor{}, and their interaction as predictor and participant as a random effect term. We employed Type III Wald chi-square tests to assess the significance of the fixed effects in the model. We corrected all post-hoc tests concerning more than two variables with the Bonferroni method.

\begin{table*}
\centering
\caption{Mean Values (M) and Standard Deviation (SD) for the \dvAnswerTime{}, \dvAccuracy{}, \dvMissedAnswerRate{} (Missed AR) of for the virtual \nback{n} task, as well as for the RTLX rating. Median and Median Absolute Deviation (MAD) for the statements "My Performance on the Virtual Task Was Very Successful" (VP) and "It was hard to Focus on the Virtual Task" (HFVT).}
\Description{A table providing an overview of the results of the virtual nback task data analysis as well as for the RTLX and the subjective ratings to the statements "My Performance on the Virtual Task Was Very Successful" (VP) and "It was hard to Focus on the Virtual Task" (HFVT).}
\resizebox{\ifdim\width>\linewidth\linewidth\else\width\fi}{!}{
\begin{tabular}{lllrrrrrrrrrrrr}
\toprule
\multicolumn{1}{c}{} & \multicolumn{1}{c}{} & \multicolumn{1}{c}{} & \multicolumn{6}{c}{Virtual Task Performance} & \multicolumn{6}{c}{Subjective Measures} \\
\cmidrule(l{3pt}r{3pt}){4-9} \cmidrule(l{3pt}r{3pt}){10-15}
\multicolumn{1}{c}{} & \multicolumn{1}{c}{} & \multicolumn{1}{c}{} & \multicolumn{2}{c}{Answer Time} & \multicolumn{2}{c}{Accuracy} & \multicolumn{2}{c}{Missed AR} & \multicolumn{2}{c}{RTLX} & \multicolumn{2}{c}{VP} & \multicolumn{2}{c}{HFVT} \\
\cmidrule(l{3pt}r{3pt}){4-5} \cmidrule(l{3pt}r{3pt}){6-7} \cmidrule(l{3pt}r{3pt}){8-9} \cmidrule(l{3pt}r{3pt}){10-11} \cmidrule(l{3pt}r{3pt}){12-13} \cmidrule(l{3pt}r{3pt}){14-15}
Walking Task & Difficulty & Anchoring & M & SD & M & SD & M & SD& M & SD & Med. & MAD & Med. & MAD\\
\midrule
No Walking & 1-back & Hand & .97 & .38 & .95 & .21 & .02 & .13 & 3.94 & 2.82 & 7.0 & .00 & 2.0 & 1.48\\
No Walking & 1-back & Head & .97 & .42 & .95 & .22 & .02 & .15& 3.02 & 2.47 & 7.0 & .00 & 1.0 & .00\\
No Walking & 1-back & Torso & .94 & .37 & .94 & .24 & .02 & .15 & 2.97 & 1.98 & 7.0 & .00 & 2.0 & 1.48\\
No Walking & 2-back & Hand & 1.27 & .53 & .87 & .33 & .03 & .16 & 5.51 & 3.36 & 6.0 & 1.48 & 2.5 & 2.22\\
No Walking & 2-back & Head & 1.18 & .46 & .85 & .36 & .07 & .25 & 5.12 & 3.20 & 6.0 & 1.48 & 2.0 & 1.48\\
No Walking & 2-back & Torso & 1.22 & .55 & .87 & .34 & .04 & .19 & 5.49 & 2.88 & 6.0 & 1.48 & 3.0 & 2.22\\
\addlinespace
Walking & 1-back & Hand & 1.16 & .42 & .90 & .30 & .02 & .15 & 6.44 & 3.58 & 6.0 & 1.48 & 5.0 & 2.97\\
Walking & 1-back & Head & 1.09 & .40 & .92 & .28 & .02 & .14 & 5.30 & 3.42 & 6.0 & 1.48 & 3.0 & 1.48\\
Walking & 1-back & Torso & 1.29 & .56 & .90 & .31 & .02 & .14 & 5.86 & 3.50 & 6.0 & 1.48 & 3.0 & 2.22\\
Walking & 2-back & Hand & 1.43 & .57 & .78 & .41 & .06 & .24 & 8.26 & 3.32 & 5.0 & 1.48 & 5.0 & 1.48\\
Walking & 2-back & Head & 1.38 & .60 & .81 & .39 & .06 & .24 & 7.71 & 3.59 & 5.5 & .74 & 5.0 & 2.97\\
Walking & 2-back & Torso & 1.60 & .68 & .76 & .43 & .09 & .28 & 8.29 & 3.68 & 5.0 & 2.22 & 3.5 & 2.22\\
\bottomrule
\end{tabular}}
\end{table*}

\subsubsection{Walking Error}
We found a higher average \dvWalkingError{} compared to the calibration walking in the baseline condition ranging from \percentVal{107.2}{71.4} for \condition{1}{\pHead{}} to \percentVal{110.8}{71.5} for \condition{1}{\pHand{}}, see \autoref{fig:results:walking:walking_error}.

The analysis indicated a significant (\chisq{2}{10.31}{<.01}) main effect of the \ivAnchor{}. Post-hoc tests only showed significant differences between \pHand{} (\percentVal{110.6}{71.1}) and \pHead{} (\percentVal{108.5}{71.7}), $p<.05$. We did not find other main or interaction effects.

\subsubsection{Stride Duration}

We analyzed the stride duration as a measure of the efficiency of our participants while walking. For all conditions, we found longer step durations compared to the baseline condition, ranging from \percentVal{114.1}{73.3} (\condition{1}{\pHead{}}) to \percentVal{124.5}{117.2} (\condition{2}{\pTorso{}}), see \autoref{fig:results:walking:stride_duration}.

The analysis indicated a significant (\chisq{1}{6.79}{<.01}) main effect of the \ivDifficulty{}. Post-hoc tests confirmed significantly higher stride durations for \nback{2} (\percentVal{120.1}{95.1}) compared to \nback{1} (\percentVal{117.9}{79.8}).

Further, we found a significant (\chisq{2}{57.53}{<.001}) main effect of the \ivAnchor{}. Post-hoc tests indicated significant differences between all groups, ranging from \percentVal{115.6}{77.7} (\pHead{}) over \percentVal{117.2}{79.6} (\pHand{}) to \percentVal{124.5}{104.2} (\pTorso{}), \pHand{} - \pHead{}: $p<0.05$, otherwise $p<.001$. We did not find an interaction effect between the factors.

\subsubsection{Stride Length}

We also analyzed the stride length as a measure of participants' efficiency while walking. We found stride length ranging from \percentVal{98.0}{0.74} (\condition{1}{\pHand{}}) to \percentVal{103.8}{77.6} (\condition{1}{\pTorso{}}), see \autoref{fig:results:walking:stride_length}.

The analysis indicated a significant (\chisq{2}{17.59}{<.001}) main effect of the \ivAnchor{}. Post-hoc tests showed a significantly longer stride length for both, \pHead{} (\percentVal{100.6}{76.1}) and \pTorso{} (\percentVal{102.5}{76.5}), compared to \pHand{} (\percentVal{98.1}{73.8}). We did not find any other main or interaction effects.

\subsubsection{Stride Width}

Further, we analyzed the stride width as another measure of the participants' efficiency while walking. Again, we found higher stride width compared to the baseline walking over all conditions, ranging from \percentVal{102.0}{35.6}  for \condition{2}{\pTorso{}} to \percentVal{107.9}{49.2} for \condition{2}{\pHead{}}, see \autoref{fig:results:walking:stride_width}.

The analysis indicated a significant (\chisq{1}{39.01}{<.001}) main effect of the \ivDifficulty{}. However, post hoc tests did not confirm significant differences between \nback{1} (\percentVal{104.5}{38.7}) and \nback{2} (\percentVal{104.1}{41.3}).

Further, the analysis indicated a significant (\chisq{2}{17.31}{<.001}) main effect of the \ivAnchor{}. Post-hoc tests confirmed significant differences between all groups with rising stride width from \pTorso{} (\percentVal{102.6}{37.3}) over \pHand{} (\percentVal{104.3}{37.5}) to \pHead{} (\percentVal{105.9}{44.5}), \pHand{} - \pHead{}: $p<.05$, otherwise $p<.001$.

Finally, the analysis indicated a significant (\chisq{2}{65.31}{<.001}) interaction effect between both factors. We found that \nback{2} resulted in a larger stride width for \pHead{} (\percentVal{107.9}{49.2} - \percentVal{103.9}{39.2}, $p<.001$). In contrast, we found that \nback{2} resulted in a \emph{smaller} stride width for \pHand{} (\percentVal{102.2}{37.1} - \percentVal{106.2}{37.9}, $<.001$) and \pTorso{} (\percentVal{102.0}{35.6} - \percentVal{103.3}{39.0}, $n.s.$).

\subsubsection{Walking Speed}

Finally, we analyzed the walking speed as a summarized measure for the walking performance. We found slower walking speeds for all conditions compared to the baseline, ranging from \percentVal{79.93}{30.00} for \condition{2}{\pTorso{}} to \percentVal{87.60}{36.45} for \condition{2}{\pHead{}}, see \autoref{fig:results:walking:speed}.

The analysis indicated no significant main (\ivDifficulty{}: \chisq{1}{.10}{>.05}, \ivAnchor{}: \chisq{2}{3.37}{>.05}) or interaction (\chisq{2}{.19}{>.05}) effects of the independent variables on the walking speed.

\subsection{Virtual Task Performance}
During the study, we logged participants' answers and answer times for the virtual task. Based on this data, we evaluate the \dvAccuracy{}, \dvAnswerTime{}, and \dvMissedAnswerRate{}.

\begin{figure}[t]
    \centering
    \includegraphics[width=\linewidth]{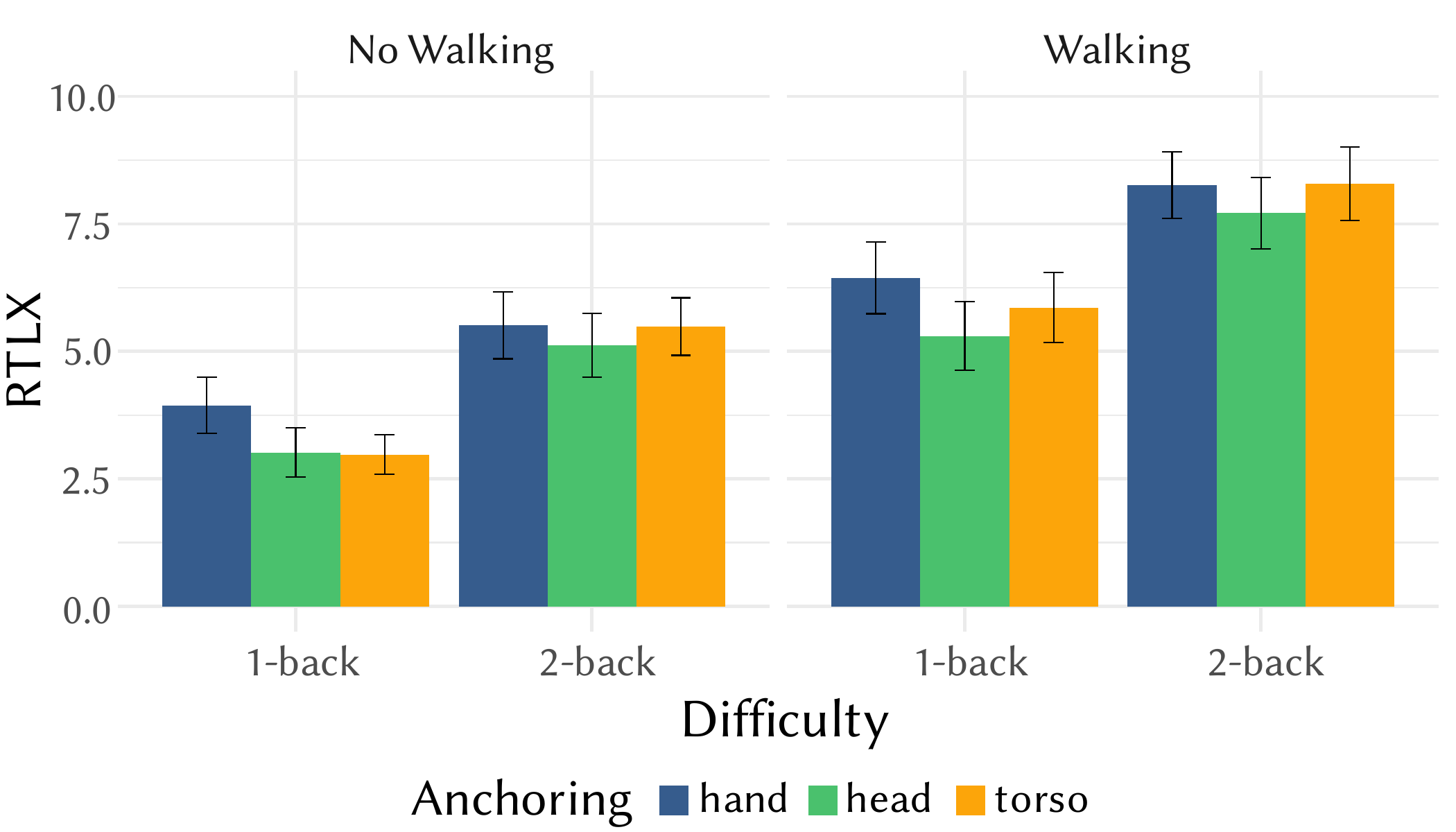}
    
    \caption{Participants RTLX Rating for each condition as an average of the scores displayed as bar charts. We display values on a y-axis 0 - 10 for better visual distinction of the differences. The RTLX results show a significant increase in perceived workload for walking compared to no walking, and for more difficult tasks (2-back) compared to easier ones (1-back). Hand anchoring consistently led to higher workload ratings, while head anchoring was associated with lower perceived workload across all conditions. Task difficulty and the physical task of walking both significantly raised participants' overall workload, with head anchoring consistently rated as less demanding than hand anchoring. There were no significant interaction effects between the variables.}
    \Description{The figure presents a bar chart comparing NASA RTLX workload scores across different AR anchoring conditions (hand, head, torso), task difficulties (1-back, 2-back), and walking versus no walking scenarios. The chart shows that workload scores are higher for walking compared to no walking, and increase with task difficulty from 1-back to 2-back. Hand anchoring consistently leads to the highest workload scores, while head anchoring results in the lowest workload across all conditions. Task difficulty and walking both significantly raise overall perceived workload.}
    \label{fig:plotTLX}
\end{figure}

\subsubsection{Answer Time}

For analyzing the \dvAnswerTime{}, we tested the data with Shapiro-Wilk's and Mauchly's tests for normality of the residuals and sphericity assumptions. We used a log-transform to correct to normality of residuals. Since sphericity was violated, we used the Greenhouse-Geisser method to correct the tests. We used three-way repeated-measures ANOVAs to identify significant effects and applied Bonferroni-corrected t-tests for post-hoc analysis. Further, we report the generalized eta-squared \ges{} as an estimate of the effect size. As suggested by \citet{bakeman_recommended_2005}, we classify these effect sizes using Cohen's suggestions~\cite{cohen_statistical_2013} as small ($>.0099$), medium ($>.0588$), or large ($>.1379$). We excluded one participant for this analysis due to technical difficulties in one condition. 

Performing a three-way RM ANOVA we found values ranging from $M= 0.94s, SD = 0.36s$ (\NoWalking{}, \nback{1}, \pTorso{}) to $M= 1.60s, SD = 0.67s$ (\Walking{}, \nback{2}, \pTorso{}). Our RM ANOVA showed a significant (\ano{1}{24}{76.48}{<.001}) main effect for the \ivPhysTask{} on the \dvAnswerTime{} with a \efETAsquared{0.17} effect size. Post-hoc tests revealed significantly  ($p<.001$) higher \dvAnswerTime{} for \Walking{} compared to \NoWalking{}. 
We also found a significant (\ano{1}{24}{74.21}{<.001}) main effect for the \ivDifficulty{} on the \dvAnswerTime{} with a \efETAsquared{0.21} effect size. Post-hoc tests revealed significantly higher \dvAnswerTime{}s for \nback{2} compared to \nback{1}. 
Further, we found a significant (\ano{1.55}{37.32}{11.09}{<.001}) main effect for the \ivAnchor{} on the \dvAnswerTime{} with a \efETAsquared{0.02} effect size. Post-hoc tests revealed significantly higher \dvAnswerTime{} for \pHand{} ($p<.01$) and \pTorso{} ($p<.001$) compared to \pHead{}. 
Finally, we could also reveal a significant (\ano{1.60}{38.48}{18.95}{<.001}) interaction effect with a \efETAsquared{0.02} effect size between \ivPhysTask{} and \ivAnchor{}. While we could not find a significant ($p > .05$) difference in the \dvAnswerTime{} between the levels of \ivAnchor{} for \NoWalking{}, for \Walking{} we found significantly higher \dvAnswerTime{}s for \pTorso{} compared to \pHand{} ($p < .01$) and \pHead{} ($p < .001$).

\begin{figure*}[th!]

	\includegraphics[width=\linewidth]{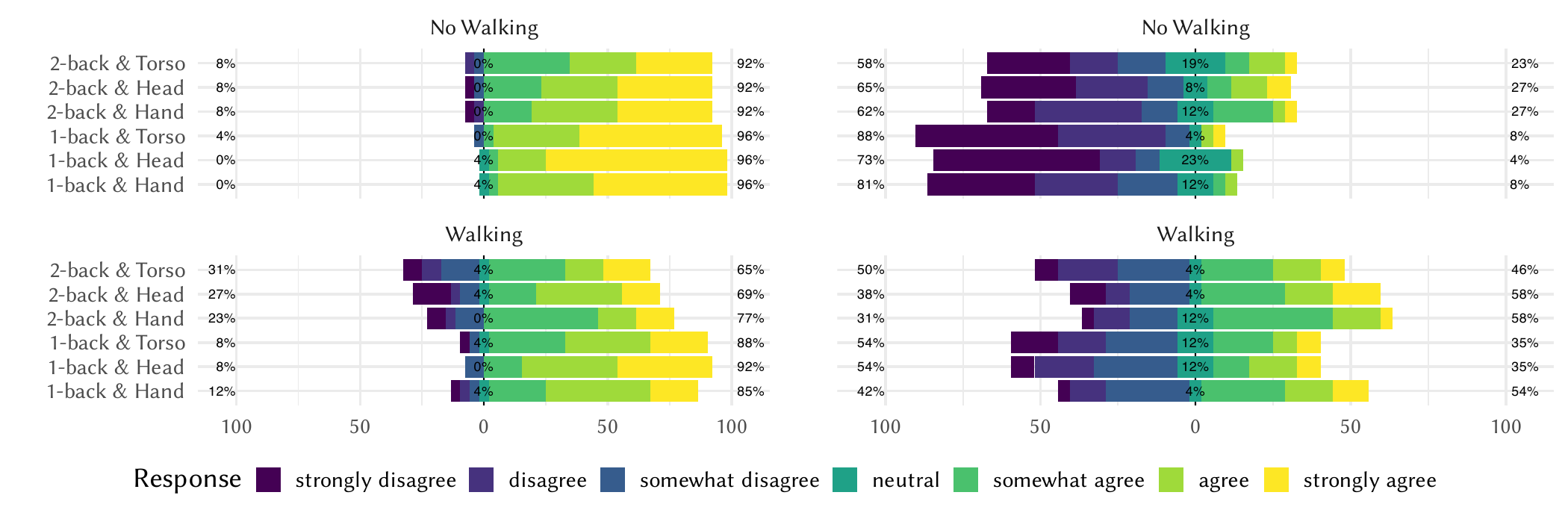}

    \vspace{-1em}

	\begin{minipage}[t]{.45\linewidth}
		\centering
	\subcaption{Virtual Performance}
    \label{fig:results:LP_VirtualPerformance}
	\end{minipage}%
    \begin{minipage}[t]{.45\linewidth}
		\centering
	\subcaption{Hard to Focus on the Virtual Task}
    \label{fig:results:LP_HardFocusVirt}
	\end{minipage}%
	\caption{Participants' ratings on a Likert scale for (a) their self reported Virtual Performance and (b) how hard they perceived it was to focus on the virtual task.}
	\Description{The figure shows two stacked bar charts comparing participants' responses on a Likert scale for self-reported virtual performance (left) and how hard it was to focus on the virtual task (right). The responses are divided by anchoring condition (hand, head, torso), task difficulty (1-back, 2-back), and walking versus no walking scenarios. In both charts, responses range from "strongly disagree" (purple) to "strongly agree" (yellow). For virtual performance, participants generally rated their performance better when not walking, with head anchoring receiving more positive ratings. In contrast, hand anchoring received more negative ratings, particularly in walking scenarios. In the second chart, focusing on the virtual task was reported as harder when walking, with hand anchoring consistently rated as more difficult than head or torso anchoring.
 }
	\label{fig:results:Lplots12_VPerf_HardFocusVirt}
\end{figure*}

\subsubsection{Accuracy}
We calculated the \dvAccuracy{} as the ratio of the correct participant answers to all virtual task trials. Missed answers were counted as "not correct", as they did not contribute to the correct answers.

We fitted a binomial generalized mixed effect model with \ivDifficulty{}, \ivAnchor{}, \ivPhysTask{}, and their interaction as predictor and participant as a random effect term. We employed Type III Wald chi-square tests to assess the significance of the fixed effects in the model. We corrected all post-hoc tests concerning more than two variables with the Bonferroni method.

The analysis indicated a significant (\chisq{1}{15.06}{<.001}) main effect of the \ivPhysTask{}. Post-hoc tests confirmed significantly higher \dvAccuracy{} for \NoWalking{} compared to \Walking{} ($p < .001$).
We further found a significant (\chisq{1}{28.50}{<.001}) main effect of the \ivDifficulty{}. Here, post-hoc tests confirmed significantly higher \dvAccuracy{} for \nback{1} compared to \nback{2} ($p < .001$).
We could not find a significant (\chisq{2}{0.73}{>.05}) main effect of the \ivAnchor{} nor interaction effects.

\subsubsection{Missed Answer Rate}
We calculated the \dvMissedAnswerRate{} as the ratio of the unanswered \nback{n} items to all virtual task trials.
Similar to the \dvAccuracy{}, we fitted a binomial model with \ivDifficulty{}, \ivAnchor{}, \ivPhysTask{}, and their interaction as predictor and participant as a random effect term.
While the analysis did not reveal a significant main effect of our independent variables, it showed significant interaction effects between them. For the interaction \ivPhysTask{}:\ivDifficulty{} we did not find significantly different \dvMissedAnswerRate{s} for the \nback{1} between the \NoWalking{} and \Walking{} conditions, but for the \nback{2} task we received significantly higher  ($p < .001$) \dvMissedAnswerRate{s} for \Walking{} compared to \NoWalking{}. For \ivDifficulty{}:\ivAnchor{} we did not find a significant different \dvMissedAnswerRate{} between the three levels of \ivAnchor{} for the \nback{1} conditions, but in the \nback{2} conditions we received significantly lower ratings for \pHand{} compared to \pHead{} ($p < .001$) and \pTorso{} ($p < .01$).

\subsection{Subjective Measures}
For the multi-factorial analysis of non-parametric data, such as the Likert questionnaires and \ac{RTLX} ratings, we performed an \ac{ART} as proposed by \citet{wobbrock_aligned_2011} and applied the ART-C procedure as proposed by \citet{elkin_aligned_2021} for post-hoc analysis.

\subsection{NASA TLX}
We calculate the \dvRTLX{} score as an average of its six subscales as suggested by \citet{hart_nasa-task_2006}.
We found a significant (\ano{1}{25}{50.16}{<.001}) main effect for the \ivPhysTask{} on participants' \ac{RTLX} ratings, with a \efETAsquared{0.66} effect size. Post-hoc tests revealed significantly ($p<.001$) higher ratings for \Walking{} compared to \NoWalking{}. 
We also found a significant (\ano{1}{25}{26.79}{<.001}) main effect for the \ivDifficulty{} on participants' \ac{RTLX} ratings, with a \efETAsquared{0.51} effect size. Post-hoc tests revealed significantly ($p<.001$) higher ratings for \nback{2} compared to \nback{1}.
We further found a significant (\ano{2}{50}{10.28}{<.001}) main effect for the \ivAnchor{} on participants' \ac{RTLX} ratings, with a \efETAsquared{0.29} effect size. Post-hoc tests revealed significantly lower ratings for \pHead{} compared to \pHand{} ($p<.05$).
We could not find significant ($p > .05$) interaction effects. \autoref{fig:plotTLX} shows the results.

\subsection{Custom Likert Questionnaire}

After each condition, participants filled out our custom Likert questionnaire. Some questions are only applicable for the \Walking{} conditions, and some for the \NoWalking{} conditions as well. The following four questions are applicable for both, and we therefore also evaluate the \ivPhysTask{} as an independent variable.

\subsubsection{My Performance on the Virtual Task Was Very Successful}
We found a significant (\ano{1}{25}{47.78}{<.001}) main effect for the \ivPhysTask{} on participants' ratings, with a \efETAsquared{0.65} effect size. Post-hoc tests revealed significantly  ($p<.001$) higher ratings for \NoWalking{} compared to \Walking{}.
We also found a significant (\ano{1}{25}{38.91}{<.001}) main effect for the \ivDifficulty{} on participants' ratings, with a \efETAsquared{0.60} effect size. Post-hoc tests revealed significantly  ($p<.001$) higher ratings for \nback{1} compared to \nback{2}.
Further, we found a significant (\ano{2}{50}{3.72}{<.05}) main effect for the \ivAnchor{} on participants' ratings, with a \efETAsquared{0.12} effect size. Post-hoc tests revealed significantly higher ratings for \pHead{} compared to \pHand{} ($p<.05$). We could not find significant ($p > .05$) interaction effects. We visualize the results in \autoref{fig:results:LP_VirtualPerformance}

\subsubsection{I Found It Hard to Focus on the Virtual Task}
We found a significant (\ano{1}{25}{36.09}{<.001}) main effect for the \ivPhysTask{} on participants' ratings, with a \efETAsquared{.59} effect size. Post-hoc tests revealed significantly  ($p<.001$) higher ratings for \Walking{} compared to \NoWalking{}.

We also found a significant (\ano{1}{25}{6.57}{<.05}) main effect for the \ivDifficulty{} on participants' ratings, with a \efETAsquared{.20} effect size. Post-hoc tests revealed significantly  ($p<.01$) higher ratings for \nback{2} compared to \nback{1}.

Further, we found a significant (\ano{2}{50}{4.09}{<.05}) main effect for the \ivAnchor{} on participants' ratings, with a \efETAsquared{0.14} effect size. Post-hoc tests revealed significantly higher ratings for \pHand{} compared to \pTorso{} ($p<.05$).

We did not find significant (all $p > .05$) interaction effects between the variables. \autoref{fig:results:LP_HardFocusVirt} shows the results.

\subsection{Custom Likert Questionnaire not considering Walking Task}
Since the following questions do not address the \NoWalking{} conditions, we did not consider \ivPhysTask{} as an independent variable for the analysis.

\begin{figure*}[th!]

	\includegraphics[width=\linewidth]{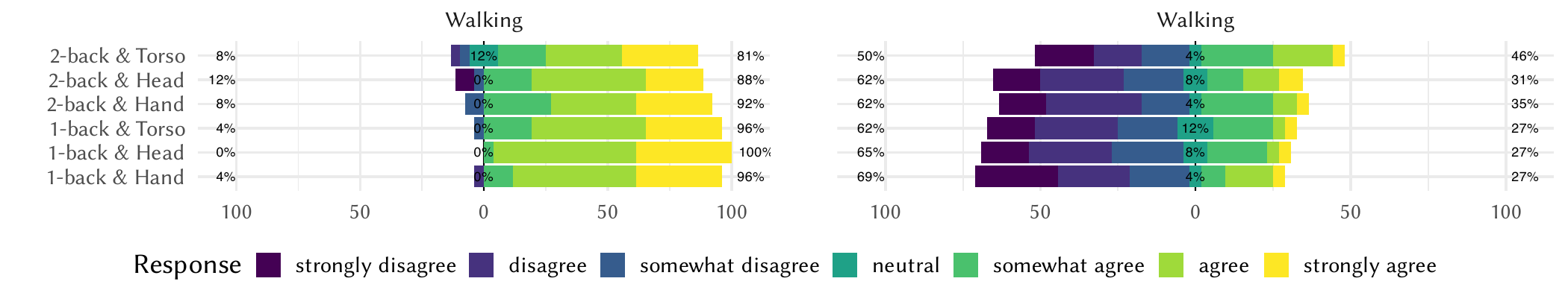}

    \vspace{-1em}

	\begin{minipage}[t]{.45\linewidth}
		\centering
	\subcaption{Physical Performance}
    \label{fig:results:LPPPerformance}
	\end{minipage}%
    \begin{minipage}[t]{.45\linewidth}
		\centering
	\subcaption{Hard to Focus on the Physical Task}
    \label{fig:results:LPHardFocusPhys}
	\end{minipage}%
 
	\caption{Participants' ratings on a Likert scale for (a) their self reported Physical Performance and (b) how hard they perceived it was to focus on the physical task.}
	\Description{A bar chart visualizing participants' Likert scale responses for two aspects: (a) self-reported Physical Performance and (b) how hard it was to focus on the physical task during walking tasks. The responses are grouped by different task conditions, including the anchoring methods (hand, head, and torso) and difficulty levels (1-back, 2-back). The color-coded bars show varying levels of agreement, from "strongly disagree" to "strongly agree." The chart indicates that participants generally found it more difficult to focus on the physical task when hand anchoring was involved, with head anchoring associated with better ratings for physical task performance.
}
	\label{fig:results:Lplots34_PPerf_HardFocusPhys}
\end{figure*}

\begin{figure*}[th!]

	\includegraphics[width=\linewidth]{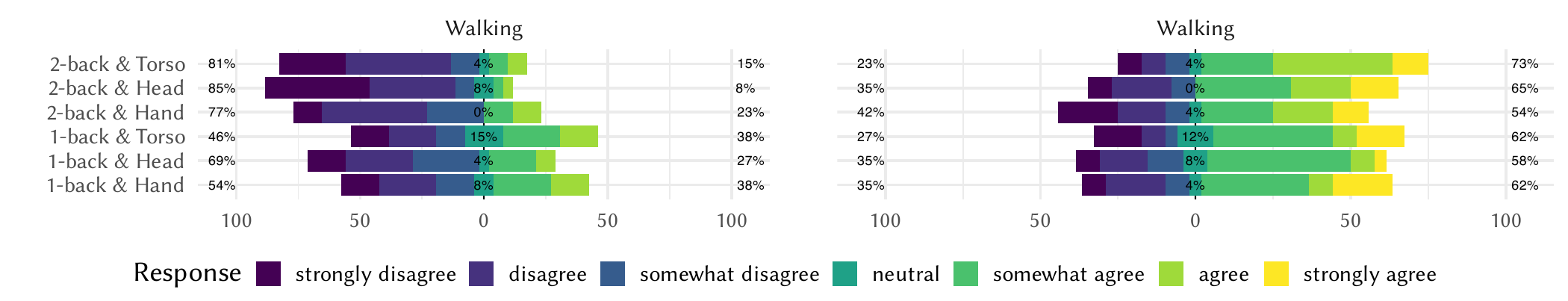}

    \vspace{-1em}

	\begin{minipage}[t]{.45\linewidth}
		\centering
	\subcaption{Physical Task more Demanding}
    \label{fig:results:LP_PhysDemand}
	\end{minipage}%
    \begin{minipage}[t]{.45\linewidth}
		\centering
	\subcaption{Virtual Task Distracted from Physical Task}
    \label{fig:results:LP_VirtDistractedPhys}
	\end{minipage}%
 
	\caption{Participants' ratings on a Likert scale regarding the (a) Physical Task was more demanding than Virtual Task and (b) Virtual Task distracted from the Physical Task}
	\Description{A bar chart showing participants' Likert scale responses on two questions: (a) whether the physical task was more demanding than the virtual task and (b) whether the virtual task distracted them from the physical task. The bars are grouped by different task conditions, including task difficulty (1-back, 2-back) and AR content anchoring (hand, head, torso). The color-coded bars range from "strongly disagree" to "strongly agree," with higher levels of agreement indicated for physical tasks being more demanding in harder conditions (2-back) and virtual tasks causing more distraction from the physical task, especially with hand anchoring.
 }
	\label{fig:results:Lplots56_PhysDemand_VirtDistractedPhys}
\end{figure*}

\subsubsection{My Performance on the Physical Task Was Very Successful}
We found a significant (\ano{1}{25}{11.27}{<.01}) main effect for the \ivDifficulty{} on participants' ratings, with a \efETAsquared{.31} effect size. Post-hoc tests revealed significantly  ($p<.01$) higher ratings for \nback{1} compared to \nback{2}.
We could not find a significant ($p > .05$) main effect for \ivAnchor{} nor interaction effects between the variables. The results are visualized in \autoref{fig:results:LPPPerformance}.

\subsubsection{I Found It Hard to Focus on the Physical Task}
We did not find any significant ($p > .05$) main effect for \ivDifficulty{} and \ivAnchor{} nor interaction effects between the variables. \autoref{fig:results:LPHardFocusPhys} shows participants' ratings.

\subsubsection{The Physical Task Was More Demanding Than the Virtual Task}
We found a significant (\ano{1}{25}{11.64}{<.01}) main effect for the \ivDifficulty{} on participants' ratings, with a \efETAsquared{0.31} effect size. Post-hoc tests revealed significantly  ($p<.001$) higher ratings for \nback{1} compared to \nback{2}.
We also found a significant (\ano{2}{50}{3.66}{<.05}) main effect for the \ivAnchor{} on participants' ratings, with a \efETAsquared{0.12} effect size. Post-hoc tests revealed significantly higher ratings for \pHand{} compared to \pHead{} ($p<.05$). We could not find any significant ($p > .05$) interaction effects.  The results are visualized in \autoref{fig:results:LP_PhysDemand}.

\subsubsection{The Virtual Task Distracted Me From the Physical Task}
We did not find any significant ($p > .05$) main effect for \ivDifficulty{} and \ivAnchor{} nor interaction effects between the variables. \autoref{fig:results:LP_VirtDistractedPhys} shows participants' ratings.

\section{Discussion}
\label{sec:discussion}

We investigated how different AR content anchoring — hand, head, and torso — impacted performance and perceived workload in a dual-task scenario combining walking and a working memory task. Overall, head anchoring supported this scenario best, with participants showing fewer walking errors, faster virtual task responses, and lower perceived workload. In contrast, hand anchoring led to slower walking and higher cognitive demands, particularly under more difficult tasks. Across all conditions, increased task difficulty worsened both virtual task performance and walking stability. Moreover, participants perceived the dual-task settings as more demanding. 
Here, we discuss our results presented in \autoref{sec:results} in light of our research questions.

\subsection{Head Anchoring Supports Virtual and Walking Task Performance but Can Increase Missed Responses in More Difficult Tasks}

In our user study, head-anchoring supported participants best in handling cognitive-motor interference, thus performing well in both, the virtual and walking task. Participants answered faster and with comparable accuracy to other anchoring methods. For the walking task, participants demonstrated smaller errors and quicker steps, reflecting accurate and efficient navigation. Additionally, participants reported lower overall task demand in the head-anchored condition. 

These results are contrary to our original expectation that anchoring to the head would lead to lower walking performance compared to the hand and torso, because of cognitive-motor interference. We reasoned that the forced display in the field of vision could direct attention to the virtual task and hinder the performance of the physical task. 
Given this, we anticipated that head anchoring, by fixing virtual content in the FOV, would increase the cognitive load on users, leading to poorer walking performance.

\begin{table*}[t!]
\centering
\caption{Median (Med.) and Median Absolute Deviation (MAD) for the statements "My Performance on the Physical Task Was Very Successful." (PP), "Hard to Focus on the Physical Task." (HFPT), "The Physical Task Was More Demanding Than the Virtual Task." (PTMD), and "The Virtual Task Distracted Me From the Physical Task" (VDPT).}
\Description{A table providing an overview of the Median and Median Absolute Deviation for four Likert statements "My Performance on the Physical Task Was Very Successful." (PP), "Hard to Focus on the Physical Task." (HFPT), "The Physical Task Was More Demanding Than the Virtual Task." (PTMD), and "The Virtual Task Distracted Me From the Physical Task" (VDPT).}
\resizebox{\ifdim\width>\linewidth\linewidth\else\width\fi}{!}{
\begin{tabular}{llrrrrrrrr}
\toprule
\multicolumn{1}{c}{} & \multicolumn{1}{c}{} & \multicolumn{2}{c}{PP} & \multicolumn{2}{c}{HFPT} & \multicolumn{2}{c}{PTMD} & \multicolumn{2}{c}{VDPT} \\
\cmidrule(l{3pt}r{3pt}){3-4} \cmidrule(l{3pt}r{3pt}){5-6} \cmidrule(l{3pt}r{3pt}){7-8} \cmidrule(l{3pt}r{3pt}){9-10}
Difficulty & Anchoring & Med. & MAD & Med. & MAD & Med. & MAD & Med. & MAD\\
\midrule
1-back & Hand & 6 & 0.74 & 2.5 & 2.22 & 3 & 2.97 & 5.0 & 2.97\\
1-back & Head & 6 & 0.00 & 3.0 & 1.48 & 3 & 1.48 & 5.0 & 1.48\\
1-back & Torso & 6 & 1.48 & 3.0 & 1.48 & 4 & 2.22 & 5.0 & 1.48\\
2-back & Hand & 6 & 1.48 & 3.0 & 2.22 & 2 & 1.48 & 5.0 & 2.97\\
2-back & Head & 6 & 1.48 & 3.0 & 1.48 & 2 & 1.48 & 5.0 & 2.22\\
2-back & Torso & 6 & 1.48 & 3.5 & 2.22 & 2 & 1.48 & 5.5 & 0.74\\
\bottomrule
\end{tabular}}
\end{table*}

In contrast, we found that head anchoring did not impair walking performance and even supported it most of all levels. We hypothesize that head anchoring reduces the need for users to adjust their FOV, allowing them to more easily focus on both tasks. This finding aligns with the work of \citet{kishishita_analysing_2014}, who demonstrated that wide FOV displays in AR minimize the need for physical adjustments, such as head movements, thus optimizing attention allocation and task efficiency. Similarly, \citet{cao_walking_2019} found that when AR content is anchored within the user’s FOV, it reduces the cognitive load associated with dividing attention between virtual and physical inputs, supporting our conclusion that head anchoring leads to improved task performance in both domains.

Our virtual task design occupied only a small portion of the participants' FOV,  with the sphere occupying approximately 16\% of the vertical FOV and 24\% of the horizontal FOV, which may also have contributed to these results. \citet{lu_glanceable_2020} showed that head-worn interfaces that position content at the periphery of vision, such as the "head-glance" interface, support unobtrusive information access. Here, our head-anchored interface, though centrally located, still allowed participants to maintain a significant portion of their peripheral vision. This likely enabled them to remain aware of the physical world while engaging with the virtual task, minimizing the need for constant visual adjustments and distractions, and helping them balance both tasks more effectively. 
As \citet{manakhov_gaze_2024} noted, larger interfaces requiring more head movements to keep the virtual and physical environments in view could lead to different outcomes. Therefore, while head anchoring proved effective in our study, larger or more complex virtual interfaces might yield different results, particularly in more demanding scenarios.

Interestingly, while head anchoring performed well across several metrics, we observed increased missed responses during more difficult tasks, particularly when participants simultaneously walked. This observation can be interpreted through Lavie’s load theory\,\citep{lavie_blinded_2014}, which posits that attentional resources become fully occupied by the primary task under high cognitive load. In this case, participants may have prioritized maintaining walking stability over engaging with the virtual task, resulting in the "filtering out" of virtual content. This selective allocation of attention under high load has been observed in other AR research, where physical navigation is often prioritized over virtual task engagement when cognitive demands are high \cite{luo_where_2022}.

Our findings contribute to existing research by demonstrating the benefits of head anchoring in mobile AR settings, particularly in dual-task scenarios where both virtual and physical tasks must be performed simultaneously. Previous studies \cite{kishishita_analysing_2014, luo_where_2022} primarily focused on AR interactions in static or simplified walking tasks. Our work expands on this by showing that head anchoring supports task performance and reduces cognitive load in dynamic conditions. However, our results also suggest that participants may shift their attention away from the virtual task in high demands. This highlights the need for further investigation into how AR head-anchored systems balance attention demands across physical and virtual tasks.

\subsection{Torso Anchoring Decreases Virtual Performance and Leads to Slower and Bigger Steps}

Torso-anchored content led to a decline in participants' virtual task performance, with slower responses and more missed answers in the difficult conditions. This likely occurred because participants had to adjust their field of view by moving their heads to alternate between the virtual task and their physical environment. 
In contrast to head anchoring, where the virtual content remains constantly in the field of vision, torso anchoring requires frequent head movements, as the virtual task and the parts of the physical world relevant to walking cannot be kept in the FOV simultaneously. This leads to inevitably missing out on events in either the physical or virtual world when they happen at the same time.

Despite the absence of significant changes in walking errors, participants walked with slower, more deliberate steps when content was torso-anchored, likely compensating for the increased cognitive load and head movements required to manage both tasks. In our study, torso anchoring introduced a cautious walking style but at the cost of efficiency, as participants had to frequently shift focus between tasks.

While torso anchoring did not significantly affect overall task demand, participants perceived it as "easier" to focus on the virtual task. This perception may stem from the fixed content position, which allowed participants to mentally "store" the task location, even though their actual performance did not reflect this ease. This fixed location required additional head movements, which likely contributed to the missed virtual task answers.

With regard to our research questions, we conclude that torso anchoring is generally the least favorable option for longer-term interaction while walking, as it does not excel in either walking or virtual performance. However, torso anchoring may still be a valuable option in specific scenarios. \citet{liu_datadancing_2023} investigated how torso-mounted AR interfaces can provide passive notifications and static data visualization, which are less demanding in terms of continuous interaction. In such cases, torso anchoring may allow users to glance at information without the need for frequent updates or immediate responses, making it suitable for slower-paced tasks where deliberate movements are acceptable. Similarly, \citet{zhou_eyes-free_2020} explored torso-mounted interfaces in collaborative tasks, where virtual content remains fixed in the environment, reducing the need for constant interaction or head movements. These scenarios suggest that torso anchoring could be effective in use cases where virtual content is supplementary to physical tasks or where users benefit from a fixed virtual reference point, as opposed to dynamic, interaction-heavy environments.

Future studies should investigate how interfaces employing torso anchoring can be optimized for specific task types, especially in situations where cognitive-motor interference is minimal or where slower movements do not hinder task performance. In particular, designing AR interfaces that reduce the need for frequent head adjustments while using torso-mounted displays could improve usability and efficiency, offering a niche solution for certain AR applications.

\subsection{Hand Anchoring Slows Down Users Virtually and Physically, Increases Demand, but Reduces Missed Answers in More Difficult Virtual Tasks}

We observed slower response times in virtual tasks for hand-anchoring, with no significant effect on accuracy. While missed answers did not differ for lower-demand tasks, hand-anchoring resulted in significantly fewer missed answers in more demanding tasks compared to the other anchoring points.

We speculate that users with hand anchoring do not necessarily prioritize the virtual task, but rather balance their attention between the virtual and physical environments. Hand anchoring allows them to adjust how much of each they see without moving their head, unlike head or torso anchoring, where the virtual content either dominates the FOV or shifts the view of the physical world. Although this balancing takes more time, slowing down response times, it helps users maintain accuracy, especially in more complex tasks, by reducing missed answers while keeping a stable interaction with both tasks.
This shift in focus likely explains fewer missed answers in high-demand tasks but also leads to slower performance. \citet{chun_real-time_2013} support this, showing that hand-based AR interactions increase completion times due to demands for executing fine-grain hand movements.

Hand anchoring resulted in more walking errors, with slower strides than head anchoring but faster than torso and smaller stride lengths, indicating reduced walking efficiency. The stride width was wider than torso but narrower than head anchoring, suggesting a more conservative walking style \cite{ko_stride_2007}. Participants took wider strides with hand anchoring in easier virtual tasks, likely due to the relative ease of switching between tasks. However, as task complexity increased, stride width narrowed, similar to torso anchoring, as participants shifted their focus to the virtual task at the expense of walking stability.

Participants also perceived hand anchoring as more demanding than head anchoring, particularly in the physical task. They reported difficulty focusing on the virtual task, especially compared to torso anchoring. This subjective report is confirmed by behavioral and gait results, where hand anchoring slowed reaction times and walking speed. The frequent switching of attention between the virtual content and the physical environment likely contributed to this increased workload, slowing participants down in both tasks and amplifying the effort required to manage the dual-task scenario.

Hand anchoring may seem less favorable due to slower performance in both tasks, but it reduced missed answers in more complex scenarios, suggesting it altered users' performance trade-off. Rather than quickly switching between tasks, users likely adopted a more conservative approach, prioritizing accuracy in the virtual task over speed, as seen in multitasking scenarios \cite{salvucci_multitasking_2005}. In fact, \citet{salvucci_multitasking_2005} highlights how users are likely to adopt a more conservative strategy when managing multitasking demands. Here, instead of switching between tasks quickly, users prioritize the virtual task for accuracy, similar to how drivers adjust their behavior to maintain safety when engaged in secondary tasks. This suggests that hand anchoring creates a mental space where users can control how and when to engage with virtual content, even at the cost of slower task completion. This trade-off, as observed in hand-anchored tasks, could be beneficial in contexts where accuracy is critical, even if it means slower overall performance.

\subsection{Influence of Virtual and Physical Task Difficulty}

Higher task difficulty led to slower response times and lower accuracy in virtual tasks across all anchorings, with no effect on missed answers. Walking performance saw minimal impact, except for slower strides under higher task demands, while subjective workload increased in line with reduced task efficiency.

These findings align with prior research on dual-task interference but also reveal mixed patterns. While prior work \cite{deblock-bellamy_virtual_2021, kao_effects_2018} demonstrated that increasing cognitive and physical task demands typically lead to prioritization of cognitive tasks over physical performance, our results suggest a different strategy. Although virtual task difficulty led to longer response times and lower accuracy, participants appeared to prioritize walking performance over virtual task engagement. This was reflected in the lack of significant differences in walking errors, stride length, or stride width but increased stride duration under higher virtual task demands. This indicates participants were slowing down to maintain stable walking, sacrificing virtual task efficiency in the process.

This behavior mirrors previous findings \cite{wollesen_differences_2019, li_evaluating_2024} that observed a cognitive-motor trade-off where participants shifted attention to maintain physical task stability when faced with complex scenarios. In our study, participants responded to increased virtual task difficulty by slowing their strides and focusing on maintaining walking performance, prioritizing locomotion over virtual task accuracy as complexity increased. 
This indicates that participants actively manage their physical performance when cognitive demand escalates, preferring to ensure stability in walking over speed or accuracy in virtual task completion.


\section{Limitations and Future Work}
\label{sec:limitions}

The results of our user study demonstrate a relationship between virtual task performance and walking efficiency, highlighting how different AR content anchors can influence both. 
However, several limitations in our methodology and technical setup highlight opportunities for further refinement, which we discuss below.

\subsection{Ecological Validity and Real-World Applicability}
One primary limitation of our study is the artificial nature of the tasks employed, which do not reflect future everyday AR use. We designed the walking task to balance realism with controlled data collection and participant safety. 
Although path-finding in the dynamically changing walking task required some attention from participants, it under-represents real-world walking in a crowded environment. This abstraction allowed participants to direct much of the cognitive effort towards the virtual task on the \ac{HMD}. The virtual task, an \nback{n} working memory task, while well-established in the literature \cite{chiossi_adapting_2023, chiossi_virtual_2022}, does not necessarily reflect the types of activities users engage in during everyday AR use. We deliberately selected these somewhat artificial tasks for the physical and virtual world in order to create a reliable foundation with high internal validity for future work.

Investigating more ecologically valid tasks could lead to a better understanding of how AR anchoring impacts cognitive-motor interference. This includes tasks such as having meetings \cite{chang_exploring_2024}, texting \cite{lu_itext_2021}, manual object manipulation\,\cite{goh_3d_2019}, or navigating through complex physical spaces while receiving AR overlays \cite{kumaran_impact_2023}. This approach could further benefit from \textit{in situ} studies. 
Further, our study's focus on a virtual task superimposed on a walking task did not require interaction with physical and virtual information. Previous research showed that integrating both types of information impacts attentional load \cite{vortmann_eeg-based_2019, chiossi_searching_2024}, thus potentially increasing the cognitive demands on users. This increased attentional burden could lead to slower task performance, higher error rates, and greater mental fatigue, especially when users frequently switch between virtual and physical stimuli.

Future research should anchor AR content \textit{in situ}, requiring users to interact with both physical and virtual elements while on the go. While this study focused on body-centric anchoring, integrating adaptive world-anchored approaches could enhance the ecological validity of mobile AR systems. Recent advancements in image segmentation and object recognition enable more accurate anchoring of AR content in the real world \cite{lages_walking_2019} or directly into objects \cite{zhou_eyes-free_2020, lan_edge-assisted_2022, han_blendmr_2023}, which can allow for world-anchored display of AR content that still moves with the users. 

Approaches such as SemanticAdapt \cite{cheng_semanticadapt_2021} and SituationAdapt \cite{li_situationadapt_2024} demonstrate how AR content can dynamically adjust to environmental and user context. In our mobile setting, these techniques could extend the dynamic walking task by incorporating environmental cues, allowing AR content to anchor onto physical objects detected along the path. For instance, AR systems could leverage image segmentation to place content on nearby landmarks or objects, dynamically adjusting its position based on proximity, object type, or user activity. This would ensure the content remains relevant and accessible while maintaining both physical navigation and task performance.

Integrating adaptive approaches into mobile AR systems could reduce user distraction and cognitive overload by aligning content placement with task requirements and environmental demands. Anchoring content onto movable objects or landmarks simulates more realistic conditions, offering deeper insights into how AR content placement affects cognitive-motor performance and usability in dynamic scenarios.

However, adaptive anchoring poses significant challenges, including maintaining stable placement during user motion, managing real-time processing demands in complex environments, and minimizing additional cognitive load. Ensuring privacy, safety, and responsiveness under hardware constraints further complicates implementation. Tackling these challenges is key to achieving seamless and effective adaptive AR interactions in dynamic environments.

\subsection{Physiological Measures for Implicit Evaluation} 
Our study used motion tracking to evaluate walking performance. The integration of additional physiological data, such as eye tracking, could shed light on how the placement of AR content affects the user's performance and cognitive load~\cite{lindlbauer_context-aware_2019} on an implicit level~\cite{kim_assessing_2024}. Along with this, analyzing the head pitch and rotation during the dual task could provide insights into user behavior and performance. While we consider this a relevant and important direction for future work, we decided against adding these dependent variables due to the additional complexity in an already complex setup.

Further, mobile EEG (mEEG) could improve multimodal evaluation by recording real-time brain activity during naturalistic tasks, as demonstrated in recent studies combining EEG with AR to measure attentional responses \cite{krugliak_towards_2022, stringfellow_recording_2024}. The combination of mEEG and AR enables the evaluation of cognitive processes like attentional shifts and workload in dynamic environments. Future research could employ these methods to investigate how different AR anchorings and varying virtual and physical information blends influence cognitive and physical performance.

\subsection{Screen Estate, Virtual Task Size, and FOV Limitations} Our study utilized relatively small virtual elements, which likely contributed to improved visibility of the real-world environment and, hence, better walking performance. However, the limited screen estate of our virtual task may not generalize to more complex AR applications that require larger FOVs. As \citet{azuma_survey_1997} noted in early work on AR, the size and placement of virtual objects within the user's FOV significantly impact both task performance and user experience.

Future research should systematically vary the size and position of AR content within the user's FOV in mobile dual-task scenarios. We aimed for a consistent size and distance of the virtual objects between the different conditions and participants, however, due to differences in body size and proportions, we acknowledge, that these were not identical in all cases. Future work should investigate these as variables in a controlled study, to investigate their influence. While in this study we focused on the anchoring of the content, also the relative position to these anchor points in form of UI placement is a relevant direction for future studies. Larger or more central AR content could obstruct the user's view of the physical world, leading to different walking behaviors and potentially more errors \cite{kruijff_influence_2019}. Similarly, content that requires users to shift their gaze between the physical and virtual environments constantly could result in greater cognitive load and motor interference. Investigating these factors in more detail, perhaps through non-transparent task canvases or virtual windows, would allow researchers to better understand the trade-offs between virtual content visibility and real-world task performance\,\cite{billinghurst_survey_2015}.

Additionally, the height placement of AR content presents a critical design consideration for mobile interactions. Higher placement (e.g., near the user's face) could reduce the need for head or gaze movement, improving reaction time for immediate tasks. However, it may obstruct peripheral vision and increase visual overload, hindering navigation in dynamic environments. Conversely, lower placement (e.g., near the waist) may enhance situational awareness by freeing up the upper FOV but could increase neck strain and delay interactions due to downward gaze shifts. These trade-offs highlight the importance of dynamic, context-aware adjustments of AR content height and size to balance cognitive load, task performance, and situational awareness in real-world mobile scenarios\,\cite{li_situationadapt_2024, cheng_semanticadapt_2021}.

\section{Conclusion}
\label{sec:conclusion}

In this paper, we systematically investigated the effects of AR anchors (hand, head, torso) and task difficulty on user performance and experience. Participants ($n=26$) engaged in a dual-task paradigm, performing a visual working memory task while walking along a dynamically changing path. We collected objective measures, including motion-tracking data for walking and logged responses for the virtual task, and subjective measures via post-condition questionnaires. Our results revealed significant differences in user performance and experience across anchor types and task difficulties.

Head anchoring supported fast and efficient navigation, as well as accurate responses in the virtual task with minimal cognitive load. However, in more demanding virtual tasks, it led to increased missed answers. Hand anchoring resulted in slower response times and reduced walking efficiency, as users frequently switched attention between the virtual task and physical environment. Nevertheless, it reduced missed answers in complex virtual tasks, making it beneficial in accuracy-critical contexts despite slower overall performance. Torso anchoring decreased virtual task performance as users had to adjust their field of view by moving their heads between tasks. While less suitable for prolonged interaction while walking, torso anchoring may be effective in scenarios requiring a stable reference point for supplementary information.
Overall, our findings emphasize the need to consider physical and cognitive factors when designing AR experiences, as different anchor types significantly influence user performance and experience in varied task scenarios.

\section{Open Science}
We encourage readers to replicate and expand upon our results by offering full access to our experimental setup and datasets. These materials are freely accessible on the Open Science Framework at \url{https://osf.io/5aqzf/}.

\begin{acks}
We thank Gaspare Pavei for sharing his expertise in the domain of biomechanics and for providing feedback on our walking task design.
Francesco Chiossi was supported by the Deutsche Forschungsgemeinschaft (DFG, German Research Foundation) within "AI Motive : Multimodal Intent Communication of Autonomous Systems"  (SCHM 1751/13-1) and  Project ID 251654672 TRR 161.\\ This work has been co-funded by the LOEWE initiative (Hesse, Germany) within the emergenCITY center [LOEWE/1/12/519/03/05.001\allowbreak(0016)/72].

\end{acks}

\bibliographystyle{ACM-Reference-Format}
\bibliography{references}

\newpage
\appendix
\section{Appendix}
\label{sec:AnalizingAnchoring}

In this appendix, we further investigate the effects of \ivAnchor{} within the different task combinations resulting from the two independent variables \ivPhysTask{} and \ivDifficulty{}. We investigate how anchoring influences both virtual and physical task performance across the four task combinations defined by the two independent variables \ivPhysTask{} and \ivDifficulty{}, specifically \NoWalking{} / \nback{1}, \NoWalking{} / \nback{2}, \Walking{},/,\nback{1}, and \Walking{} / \nback{2}.

After we prepared the data as described in \autoref{sec:results} we filter the data into separate data frames, reflecting the four task combinations resulting from the two IVs \ivPhysTask{} and \ivDifficulty{}. We then treat \ivAnchor{} as the only IV to study the differences in the four different task combinations.
For all dependent variables, we computed linear mixed-effects models (LMEs) with \ivAnchor{} as predictor and participant as a random effect term. We employed Type III Wald chi-square tests to assess the significance of the fixed effects in the model. We corrected all post-hoc tests with the Bonferroni method.

\begin{figure*}[ht!]

	\includegraphics[width=\linewidth]{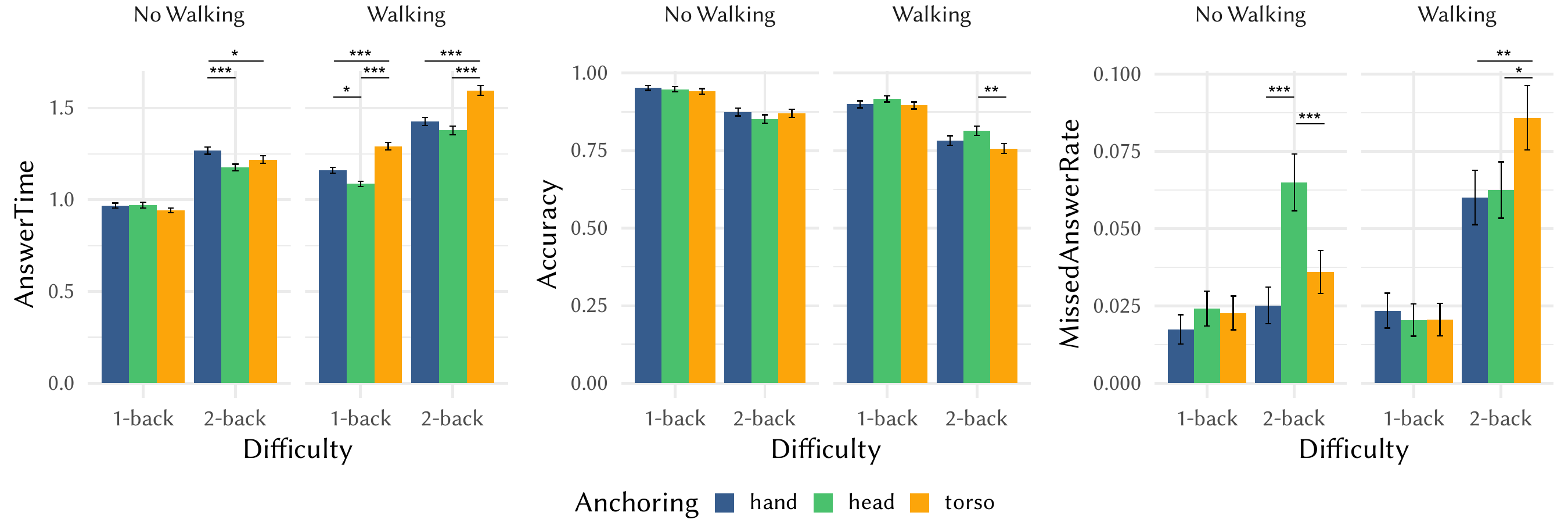}
	\begin{minipage}[t]{.3\linewidth}
		\centering
        \vspace{-1em}
		\subcaption{\dvAnswerTime{}}
        \label{fig:appendix:results:AnswerTime}
	\end{minipage}%
    \begin{minipage}[t]{.3\linewidth}
		\centering
        \vspace{-1em}
	    \subcaption{\dvAccuracy{}}
        \label{fig:appendix:results:CorrectAnswerRate}
	\end{minipage}%
    \begin{minipage}[t]{.3\linewidth}
	   \centering
       \vspace{-1em}
	   \subcaption{\dvMissedAnswerRate{}}
        \label{fig:appendix:results:MissedAnswerRate}
	\end{minipage}%
	\caption{The mean results for (a) \dvAnswerTime{} and (b) \dvAccuracy{} and (c) \dvMissedAnswerRate{} as a bar chart plot. The error bars indicate the standard error. Answer times increased when the n-back level was higher, particularly in walking conditions, and were slowest for torso anchoring. Accuracy declined slightly with task difficulty and when walking, but there was no significant effect from anchoring. Missed answer rates increased substantially with higher task difficulty, especially during walking, with hand anchoring resulting in fewer missed answers compared to head and torso anchoring in more difficult tasks. The stars symbolize significant differences (* for p < .05, ** for p < .01, and *** for p < .001) in the pairwise comparison of the levels of \ivAnchor{} within the task combinations. The results of this additional analysis are shown in \autoref{table:appendix:nback}.}
	\Description{The figure presents three bar charts illustrating the impact of AR content anchoring (hand, head, torso) and task difficulty (1-back, 2-back) on virtual task performance, with and without walking. The first chart shows "Answer Time," where response times increase with higher task difficulty and walking, and are slowest for torso anchoring. The second chart shows "Accuracy," where performance slightly decreases with task difficulty and walking, but anchoring has no significant effect. The third chart shows "Missed Answer Rate," which increases with higher task difficulty and walking, with fewer missed answers for hand anchoring compared to head and torso anchoring in the most challenging conditions.}
	\label{fig:appendix:plot:nback}
\end{figure*}

\begin{figure*}[t!]
	\includegraphics[width=\linewidth]{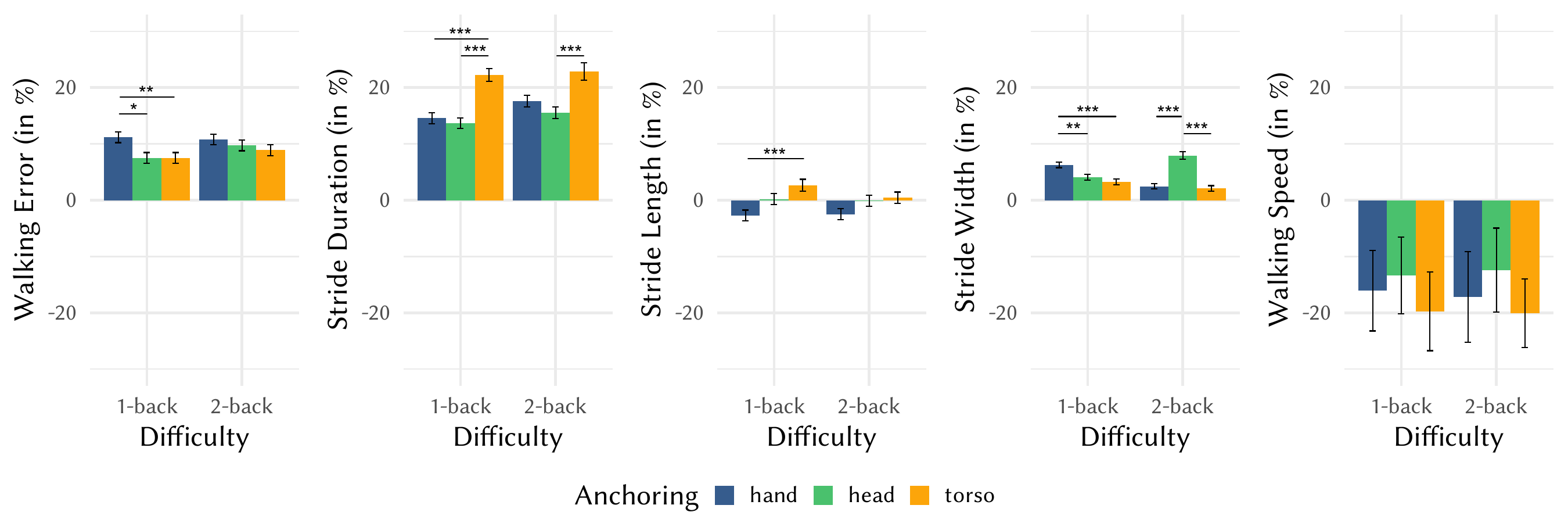}

	\begin{minipage}[t]{.2\linewidth}
		\centering  
        \vspace{-1em}
		\subcaption{\dvWalkingError{}}\label{fig:results:walking:walking_error}
	\end{minipage}%
    \begin{minipage}[t]{.2\linewidth}
		\centering
        \vspace{-1em}
		\subcaption{\dvStrideDuration{}}\label{fig:results:walking:stride_duration}
	\end{minipage}%
     \begin{minipage}[t]{.2\linewidth}
		\centering
        \vspace{-1em}
		\subcaption{\dvStrideLength{}}\label{fig:results:walking:stride_length}
	\end{minipage}%
	\begin{minipage}[t]{.2\linewidth}
		\centering
        \vspace{-1em}
		\subcaption{\dvStrideWidth{}}\label{fig:results:walking:stride_width}
	\end{minipage}%
	\begin{minipage}[t]{.2\linewidth}
		\centering
        \vspace{-1em}
		\subcaption{\dvWalkingSpeed{}}\label{fig:results:walking:speed}
	\end{minipage}%
	\caption{The dependent variables for the walking performance. All measurements are normalized using the baseline walking condition of the respective participant and, thus, represent the difference to the normal walking in percent. All error bars depict the standard error. (a) The walking error slightly increased for hand anchoring across all n-back levels. (b) Stride duration was significantly longer for torso anchoring, particularly with the 2-back task. (c) Stride length was shorter for hand anchoring and increased for head and torso anchoring, suggesting hand anchoring induces more cautious walking patterns. (d) Stride width increased with head anchoring under more difficult tasks, reflecting a potential need for greater stability when attention is focused on the virtual task. (e) Walking speed did not change significantly between the conditions. The stars symbolize significant differences (* for p < .05, ** for p < .01, and *** for p < .001) in the pairwise comparison of the levels of \ivAnchor{} within the task combinations. The results of this additional analysis are shown in \autoref{table:appendix:results:walking_withStars}.}
	\label{fig:appendix:plot:walking}
 \Description{The figure consists of five bar charts illustrating the impact of AR content anchoring (hand, head, torso) and task difficulty (1-back, 2-back) on various walking performance metrics. Chart (a) "Walking Error" shows slight increases in error for hand anchoring compared to head and torso across both task difficulties. Chart (b) "Stride Duration" shows that stride duration increases with task difficulty, with torso anchoring showing the longest durations. Chart (c) "Stride Length" shows that hand anchoring leads to shorter strides, while head and torso anchoring result in longer strides. Chart (d) "Stride Width" shows wider strides with head anchoring, particularly in more difficult tasks, indicating a need for greater stability. Chart  (e) Walking speed did not change significantly between the conditions.}
\end{figure*}

In the following we present the results for the virtual task performance in \autoref{table:appendix:nback} and the walking task performance in \autoref{table:appendix:results:walking_withStars}. We visualize the results in \autoref{fig:appendix:plot:nback} and \autoref{fig:appendix:plot:walking}, originally introduced in the \autoref{sec:results} section of this paper, now showing the significant effects  (* for p < .05, ** for p < .01, and *** for p < .001) for the comparison within the task combinations.

\newpage

\begin{table*}[htbp]
\centering
\caption{Results of the pairwise comparison of the levels of \ivAnchor{} within the task combinations for the virtual task performance reflected by Answer Time, Accuracy, and Missed Answer Rate. Significant results (p < .05) are shown in bold text.}
\begin{tabular}{llcccccc}
\toprule
Task Combination & Contrast & \multicolumn{2}{c}{Answer Time} & \multicolumn{2}{c}{Accuracy} & \multicolumn{2}{c}{Missed Answer Rate} \\
\cmidrule(lr){3-4} \cmidrule(lr){5-6} \cmidrule(lr){7-8}
& & t-ratio & p-value & t-ratio & p-value & t-ratio & p-value \\
\midrule
\textbf{No Walking - 1-back} & Hand - Head &  -0.087 & 1.000 &  0.194 & 1.000 &  -0.933 & 1.000 \\
& Hand - Torso &  1.340 & .541 &  .801 & 1.000 &  -0.682 & 1.000 \\
& Head - Torso &  1.424 & .463 &  .606 & 1.000 &  .252 & 1.000 \\
\textbf{No Walking - 2-back} & Hand - Head &  4.215 & \textbf{< .001} &  1.414 & .472 &  -5.270 & \textbf{ < .001} \\
& Hand - Torso &  2.572 &\textbf{ .031} &  .156 & 1.000 &  -1.414 & .473 \\
& Head - Torso &  -1.672 & .284 &  -1.260 & .623 &  3.863 & \textbf{< .001} \\
\textbf{Walking - 1-back} & Hand - Head &  3.183 &\textbf{ .004} &  -1.241 & .645 &  0.482 & 1.000 \\
& Hand - Torso &  -5.691 & \textbf{< .001} &  0.237 & 1.000 &  .479 & 1.000 \\
& Head - Torso &  -8.902 & \textbf{< .001} &  1.48 & .417 &  -.002 & 1.000 \\
\textbf{Walking - 2-back} & Hand - Head &  1.254 & .630 &  -1.250 & .635 &  -0.444 & 1.000 \\
& Hand - Torso &  -6.512 & \textbf{< .001} &  1.959 & .151 &  -3.141 & \textbf{.005} \\
& Head - Torso &  -7.696 & \textbf{< .001} &  3.192 & \textbf{.004} &  -2.675 & \textbf{.023} \\
\bottomrule
\end{tabular}
\label{table:appendix:nback}
\end{table*}

\clearpage

\begin{table*}[tbp]
\centering
\caption{Results of the pairwise comparison of the levels of \ivAnchor{} within the task combinations for the physical task performance reflected by the Walking Error, Stride Duration, Stride Length, Stride Width, and Walking Speed. Significant results (p < .05) are shown in bold text.}
\begin{tabular}{llcccccccccc}
\hline
Task & Contrast & \multicolumn{2}{c}{Walking Error} & \multicolumn{2}{c}{Stride Duration} & \multicolumn{2}{c}{Stride Length} & \multicolumn{2}{c}{Stride Width} & \multicolumn{2}{c}{Walking Speed} \\
\cmidrule(lr){3-4} \cmidrule(lr){5-6} \cmidrule(lr){7-8} \cmidrule(lr){9-10} \cmidrule(lr){11-12}
& & z-ratio & p-value & z-ratio & p-value & z-ratio & p-value & z-ratio & p-value & t-ratio & p-value \\
\hline
1-Back & Hand - Head &  3.062 & \textbf{.007}&  1.977 & .144 &  -2.061 & .118 &  3.201 & \textbf{.004} &  -1.440 & .470 \\
& Hand - Torso &  2.693 & \textbf{.021} &  -5.113 & \textbf{< .001} &  -3.733 & \textbf{< .001} &  4.405 & \textbf{< .001} &  .867 & 1.000 \\
& Head - Torso &  - .306 & 1.000 &  -7.065 & \textbf{< .001} &  -1.719 & .257 &  1.274 & .608 &  2.308 & .0770 \\
2-Back & Hand - Head &  .536 & 1.000 &  1.700 & .267 &  -1.787 & .222 &  -7.730 & \textbf{< .001} &  -1.440 & .470 \\
& Hand - Torso &  1.050 & .881 &  -2.386 & .051 &  -1.674 & .283 &  -0.035 & 1.000 &  .867 & 1.000 \\
& Head - Torso &  .524 & 1.000 &  -4.073 & \textbf{< .001} & .087 & 1.000 & 7.602 & \textbf{< .001} &  2.308 & .077 \\
\hline
\end{tabular}
\label{table:appendix:results:walking_withStars}
\end{table*}

\end{document}